\documentclass[11pt]{article}
\usepackage{epsfig}
\usepackage{amsfonts}
\usepackage{amssymb}
\usepackage{amsmath}
\usepackage{amsthm}
\usepackage{latexsym}
\usepackage{verbatim}
\usepackage{graphicx}
\usepackage{slashed}
\usepackage{textcomp}
\usepackage{authblk}
\usepackage{color}

\title{\bf Nonuniversality of indirect CP asymmetries in 
$D \to \pi\pi, KK$ decays}

\author[1]{Amol Dighe \thanks{amol@theory.tifr.res.in}}
\author[2]{Diptimoy Ghosh \thanks{diptimoy.ghosh@roma1.infn.it}}
\author[1]{Bhavik Kodrani\thanks{bhavik@theory.tifr.res.in}}

\affil[1]{Tata Institute of Fundamental Research,
Mumbai 400005, India}
\affil[2]{INFN, Sezione di Roma,
Piazzale A. Moro 2, I-00185, Roma, Italy}

\date{}

\newcommand{\ddbar}{D-\bar{D}}
\newcommand{\kk}{K^+ K^-}
\newcommand{\pp}{\pi^+ \pi^-}

\newcommand{\acp}{A_{\rm CP}}
\newcommand{\acpdir}{A_{\rm CP}^{\rm dir}}
\newcommand{\acpindir}{A_{\rm CP}^{\rm indir}}
\newcommand{\ycp}{y_{\rm CP}}
\newcommand{\agamma}{A_{\Gamma}}
\newcommand{\fcp}{{f_{\rm CP}}}

\def\beq{\begin{equation}}
\def\eeq{\end{equation}}
\def\barr{\begin{eqnarray}}
\def\earr{\end{eqnarray}}

\begin{document}

\maketitle

\begin{abstract}
We point out that, if the direct CP asymmetries in the
$D \to \pi^+ \pi^-$ and $D \to K^+ K^-$ decays are unequal, the indirect 
CP asymmetries as measured in these modes are necessarily unequal.
This nonuniversality of indirect CP asymmetries can be 
significant with the right amount of new physics contributions,
a scenario that may be fine-tuned, but is still viable.
A model-independent fit to the current data allows different 
indirect CP asymmetries in the above two decays.
This could even be contributing to the apparent tension between the
difference CP asymmetries $\Delta A_{\rm CP}$ measured through the
pion-tagged and muon-tagged data samples at the LHCb.
This also implies that the measurements of $A_\Gamma$ and $y_{\rm CP}$ in 
the $\pi^+ \pi^-$ and $K^+ K^-$ decay modes can be different,
and averaging over these two modes should be avoided.
In any case, the complete analysis of CP violation measurements 
in the $D$ meson sector needs to take into account the possibility 
of different indirect CP asymmetries in the $\pi^+\pi^-$ and 
$K^+ K^-$ channels.

\end{abstract}

\newpage

\section{Introduction}

The study of charge-parity (CP) violation in decays of $K$ and $B$ mesons 
have yielded path-breaking results over the past half a century. Through these
measurements, the Kobayashi-Maskawa paradigm of CP violation has been
tested from many directions, and has emerged vindicated so far. These
tests act as indirect probes of new physics beyond the Standard Model, 
and so far have not yielded any conclusive evidence for a deviation
from the Standard Model (SM) predictions. A positive identification of
deviations from the SM is often limited by the uncertainties in the
SM predictions themselves, especially in the processes that involve
decays of mesons, due to hadronic uncertainties.

The measurements of $\ddbar$ mixing and the CP violation in $D$ decays
have started yielding interesting results only in the past decade. 
One of the reasons for the $D$ decays to have come to the forefront 
so late is that the mixing as well as CP violation in the $D$ sector 
is expected to be small. In the SM, the contribution to the $\ddbar$
mixing box diagram is suppressed --- for intermediate $u$ and $c$ quarks,
due to their small masses, and for an intermediate $b$ quark, due to the
small Cabibbo-Kobayashi-Maskawa (CKM) matrix elements. 
This leads to very small values for 
both the dispersive as well as absorptive parts of the $\ddbar$ mixing 
amplitude. Indeed, the measurements give $x \equiv \Delta m / \Gamma$
and $y \equiv \Delta \Gamma / (2 \Gamma)$ to be less than ${\cal O}(1\%)$
\cite{belle-ddbar-07,babar-ddbar-10,hfag}.
Moreover, since the phases of the relevant CKM 
matrix elements are very small, the CP violation is also expected to be
not more than ${\cal O}(0.1\%)$ \cite{nir-review}.
 
The SM calculations of the mixing and CP asymmetries in the neutral $D$ 
meson system are difficult due to the mass of the charm quark --- 
it is not light enough to enable the use of the chiral perturbation theory, 
and not heavy enough to guarantee convergence of the $(1/m_c)$ expansion
in the heavy quark effective theory.
Moreover, the nonperturbative long-distance contributions to the mixing 
as well as decay amplitudes may be dominant --- since the strong coupling is 
not very small at the scale of the $D$ meson mass, and the short-distance
contributions do not have the benefit of an intermediate $t$ quark as in
the case of $B$ mesons. 
Therefore, $D$ decays are not a good place for precision measurements
of the SM parameters. However, they can still be used as probes of new
physics (NP), if the NP effects can be large compared to the SM ones
\cite{nir-review,pdg-2012}. 
Modes like $D/\bar{D} \to \pp$ and $D/\bar{D} \to \kk$ can be sensitive to 
the presence of such NP \cite{grossman-kagan-nir,kagan-sokoloff}.

While the measurements of the mixing parameters $x$ and $y$ are 
consistent with the SM estimates of $x,y \sim {\cal O}(1\%)$
\cite{fglp-02,fglnp-04}, recent measurements of CP-violating 
quantities have given us a reason to consider the presence of NP
contributions. The CP violation in $D \to \pp$ and $D \to \kk$ decays 
was constrained by E791 \cite{e791-acp-97}, FOCUS \cite{focus-acp-00},
CLEO \cite{cleo-acp-01}, and the $B$ factories 
\cite{babar-acp-08,belle-acp-12}. Recently CDF \cite{cdf-acp-11,cdf-acp-12}
and LHCb \cite{lhcb-acp-11,lhcb-acp-13-pi,lhcb-acp-13-mu} 
presented the measurements for the ``difference CP asymmetry'' $\Delta \acp$, 
the difference between the CP asymmetries in the above two decay modes, 
that was expected to cancel out some of the systematic uncertainties.
These results, obtained using the pion-tagged samples, indicated a
value of ${\cal O}(0.5\%)$ for $\Delta \acp$.
These measurements disfavored a vanishing CP asymmetry, and were also 
away from the SM prediction (leading order in $1/m_c$ 
\cite{Brod:2011re}) of ${\cal O}(0.05\%-0.1\%)$ by more than $\sim 2\sigma$.
Moreover, the latest LHCb results with the pion-tagged sample
\cite{lhcb-acp-13-pi} and the muon-tagged sample 
\cite{lhcb-acp-13-mu} have central values with opposite signs, and 
differ by $\sim 2.2 \sigma$ from each other.
The average of these two LHCb measurements has also been
recently reported \cite{lhcb-eps13}, which is consistent with
vanishing $\Delta\acp$. (However as we shall point out in this paper,
such an average need not be the right observable to look for.)

Several attempts have been made to check if the observed large CP asymmetries 
can be accommodated within the SM. 
It has been claimed that QCD penguin operators, with large strong phases,
may give rise to a significant enhancement  
\cite{Brod:2011re,Cheng:2012wr,Bhattacharya:2012ah,Li:2012cfa}.
The breaking of $SU(3)$ \cite{Pirtskhalava:2011va}, 
or $U$-spin \cite{Brod:2012ud,Feldmann:2012js} symmetries,
or of the naive $1/N_c$ counting \cite{Franco:2012ck} 
may also be a reason for the observed large $\Delta \acp$. 
However the jury is still out on whether these contributions can
account for the data without the need to go beyond the SM.

Specific NP models that can enhance the CP asymmetries 
have also been extensively studied. These include the fourth quark 
generation \cite{Rozanov:2011gj,Feldmann:2012js},
supersymmetric gluino-squark loops \cite{grossman-kagan-nir}, 
littlest Higgs model with T-parity \cite{bigi-blanke,Paul:2011ar},
flavor violation in the up sector \cite{Wang:2011uu,Bai:2013ooa}, 
models with a color-sextet diquark \cite{Chen:2012am},
models giving rise to the $t$-channel exchange of a weak doublet 
with a special flavor structure \cite{Hochberg:2011ru}, the
nonmanifest left-right symmetric model \cite{Chen:2012usa},
or models with warped extra dimensions \cite{Paul:2012ab}.
A survey of the effect of NP models that may contribute to the
difference CP asymmetry has been performed in Ref.~\cite{Altmannshofer:2012ur}.
It points out that the CP violation may be generated at the tree
level with models that involve flavor-changing couplings of $Z,Z'$ bosons, 
new charged gauge bosons, flavor-changing heavy gluon, scalar octets,
a scalar diquark, or a two-Higgs doublet with minimal flavor violation.
Models with GIM-unsuppressed fermion and scalar loops, or those with
chirally enhanced magnetic penguin operators, can also contribute to the
CP asymmetry at the loop level. 
It has been observed \cite{Giudice:2012qq} that NP models in which the 
primary source of flavor violation is linked to the breaking of chiral 
symmetry are natural candidates to explain the CP asymmetries, via enhanced 
chromomagnetic operators.
Many of these models also affect the measurements of other $D$ decay
channels, as well as the $\ddbar$ mixing and $\epsilon'/\epsilon$
in the $K$ sector, and hence the masses of new particles and couplings
in most of these models are severely constrained \cite{Isidori:2011qw}.

Identifying whether the enhancement in the CP violation in
$D \to \pp, \kk$ is from the SM or NP
is not straightforward; however, some information may be obtained from 
related decay modes. It was pointed out in Ref.~\cite{Atwood:2012ac} that, 
since the enhancement due to nonperturbative physics should only affect 
exclusive modes, an enhancement in the inclusive 
modes will point definitively to NP. 
One could also look at modes related to $\pp,\kk$ by isospin symmetry,
since this symmetry is not expected to be broken significantly.
Such a comparison could distinguish between a large penguin amplitude and
an enhanced chromomagnetic dipole operator, for example 
Ref.~\cite{Cheng:2012xb}.

The aim of this paper is not to check whether the SM or any specific NP
model explains the data. Rather, we choose to take the data at face value,
and learn in a model-independent way what they tell us about the CP violation
in the $\ddbar$ mixing and decay. To this end, we perform a fit to the data
with four complex parameters, $M_{12}, \Gamma_{12}, R_\pi$ and $R_K$. 
Here, the mixing parameters $M_{12}$ and $\Gamma_{12}$ are the 
complex-valued dispersive and absorptive components, respectively, of 
the effective $\ddbar$ mixing Hamiltonian. The other two parameters,
$$R_\pi \equiv \frac{A(\bar{D} \to \pp)}{A(D \to \pp)} \quad {\rm and} \quad 
R_K \equiv \frac{A(\bar{D} \to \kk)}{A(D \to \kk)} \; ,$$ 
are the ratios of decay 
amplitudes of a pure $\bar{D}$ and $D$ to the CP eigenstates.

The data on $\ddbar$ mixing and CP asymmetries in the $\pp$ and $\kk$
channels form the main input for the fit.
The ingredients for the fit also include the asymmetries 
$\agamma(\pi)$, $\agamma(K)$, $\ycp(\pi)$, and $\ycp(K)$, 
constructed from the ratios of effective lifetimes 
measured in the CP-eigenstate modes $D/\bar{D} \to \pp, \kk$, and  the 
(almost-)flavor-specific modes $D \to K^+ \pi^-, \bar{D} \to K^- \pi^+$
\cite{Bergmann:2000id},
which have been reported by FOCUS \cite{focus-ycp-00},
CLEO \cite{cleo-acp-01}, Belle \cite{belle-ycp-07}, 
Babar \cite{babar-ycp-08,babar-ycp-09}, and 
LHCb \cite{lhcb-ycp-11}.  
While the measurements of $\agamma(\pi)$, $\agamma(K)$ and $\ycp(\pi)$
are consistent with zero to within $2\sigma$, the asymmetry $\ycp(K)$
has been found to be nonzero to more than $4\sigma$
\cite{belle-ycp-07,babar-ycp-08,babar-ycp-09}. 
More recently, Belle \cite{belle-ycp-12} and Babar \cite{babar-ycp-12}
have reported values of $\agamma$ and $\ycp$ averaged over the 
$\pp$ and $\kk$ samples, where the $\ycp({\rm avg})$ has been
found to be nonzero to more than $\sim 4 \sigma$ in each experiment.
These asymmetries may be represented in terms of the combinations of 
the same parameters
considered above, hence the information content in these measurements
is also relevant in determining the favored parameter values,
and in fact, is commonly used \cite{hfag}.

The classification of CP violation in neutral meson systems is 
normally described in two languages.
One may talk in terms of CP violation in only mixing (deviation of
$|q/p|$ from unity), in only decay (deviation of $|R_f|$ 
from unity, where $R_f \equiv \bar{A}_f/A_f$), 
and in the interference of mixing and decay (imaginary part
of $\lambda_f \equiv (q/p)R_f$). This is the standard notation
used in the discussion of $B$ decays. 
On the other hand, one may use the language of direct vs. indirect CP
violation, which has its origins in the analyses of $K$ decays.
While we personally prefer the former formulation due to its clarity
in distinguishing the source of the CP violation, the latter one
has been used in most of the literature on the CP asymmetries in 
$D$ decays that is the focus of this paper. 
Indeed, the recent experimental data
\cite{cdf-acp-11,cdf-acp-12,lhcb-acp-11,lhcb-acp-13-pi,lhcb-acp-13-mu} 
have been interpreted in terms of the direct and indirect CP asymmetries 
($\acpdir$ and $\acpindir$, respectively) in the $\pi\pi$ and $KK$ decays.
We therefore shall refer to both the notations, at the risk of some repetition
in presenting our results and interpretations.

The interpretation of $\Delta \acp$ in terms of its direct and indirect 
components often \cite{cdf-acp-12,lhcb-acp-11,lhcb-acp-13-pi,lhcb-acp-13-mu}
takes  $\acpindir(\pi) = \acpindir(K)$.
We point out that if this condition were strictly valid, it would 
also imply $\acpdir(\pi)=\acpdir(K)$, independent of the origin
of the CP asymmetry. 
This is clearly not the case, even in the limit of 
flavor $SU(3)$ where these two quantities have the same magnitudes
but opposite signs.
Therefore, the assumption of exactly equal $\acpindir$ in the
$\pi^+\pi^-$ and $K^+K^-$ channels is, strictly speaking, not accurate.
In practice, with certain ``natural'' expectations about the 
amplitudes and phases of NP contributions, the nonuniversality of 
$\acpindir$ may turn out to be so small that it may be neglected
\cite{grossman-kagan-nir,kagan-sokoloff}, since the difference 
$\acpdir(\pi)-\acpdir(K)$ is less than ${\cal O}(0.01)$ and is expected to
contribute to the nonuniversality $\acpindir(\pi) = \acpindir(K)$
only to the second order. 
However, while searching for physics beyond the SM, the analysis of data 
should be performed without prejudice to theoretical expectations,
and alternative scenarios, however unlikely they may seem, should be 
considered.
We therefore reanalyze the current data without the approximation
$\acpindir(\pi) = \acpindir(K)$.
Our fit, in fact, shows that the preferred parameter space
allows significantly different values for $\acpindir$ in $\pp$
and $\kk$ decays. Such a difference could also contribute to 
the seemingly different $\Delta A_{\rm CP}$ values measured through the
pion-tagged and muon-tagged data samples at the LHCb
\cite{lhcb-acp-13-pi,lhcb-acp-13-mu}. 
That this nonuniversality also leads to the nonuniversality of
$\agamma$ and $\ycp$ has been indirectly alluded to in 
Ref.~\cite{gersabeck-12}.

Our paper is organized as follows. In Sec.~\ref{analytic}, we
present the analytical expressions for the time-dependent CP asymmetries
and their direct and indirect components, $\acpdir$ and $\acpindir$,
inferred from the data. We also relate $\agamma$ and $\ycp$, the quantities
obtained from the measurements of effective $D$ decay rates in different
channels, to the relevant CP-violating quantities.
In Sec.~\ref{numerical}, we perform a $\chi^2$ fit to the data and obtain
the favored values for the parameters of interest.
Section~\ref{implications} is devoted to the feasibility and 
implications of a significant nonuniversality of $\acpindir$.
Section~\ref{concl} summarizes our results and recommends taking the possible
nonuniversality in $\acpindir$ into account for future analyses of 
neutral $D$ decay data.

\section{$\ddbar$ mixing and decay: formalism}
\label{analytic}

We follow the analysis of $\ddbar$ mixing and decay as in 
Refs.~\cite{nir-review,gersabeck-12}. Since the notations vary from analysis
to analysis, we repeat the relevant steps to clarify our notation.
In the $(D, \bar{D})$ flavor basis, the effective Hamiltonian 
$H = M - i\Gamma/2$ is not diagonal. The off-diagonal elements of 
the dispersive and absorptive components, i.e. $M_{12}$ and $\Gamma_{12}$,
are responsible for the $\ddbar$ mixing. 
The mass eigenstates are given by
\begin{equation}\label{defn_mass_eigen_state}
|D_L \rangle =  p |D \rangle + q |\bar{D} \rangle \;, \quad 
|D_H \rangle =  p |D \rangle - q |\bar{D} \rangle \; ,
\end{equation}
where
\beq
|q|^2 + |p|^2 = 1 \; , \quad
\left( \frac{q}{p} \right)^2 = \frac{M_{12}^* - \frac{i}{2} \Gamma_{12}^*}{
M_{12} - \frac{i}{2} \Gamma_{12}} \; .
\eeq
The deviation of $|q/p|$ from unity corresponds to CP violation in mixing.
Note that as opposed to the mixing in the neutral $B$ systems
($B-\bar{B}, B_s - \bar{B}_s$) where $|\Gamma_{12}| \ll |M_{12}|$, 
here we have the possibility of $\Gamma_{12}$ being of the same
order as $M_{12}$ or even a few times larger \cite{Bergmann:2000id}. 
If in addition, $M_{12}$ and $\Gamma_{12}$ have significantly different
phases, then  $|q/p|$ can differ substantially from unity, and the effects 
of this CP violation in mixing need to be taken care of in the analysis.

With the mass difference and the decay width difference of the
interaction eigenstates $D_{H,L}$ defined as
\begin{equation}
\Delta m = m_H - m_L \; , \quad \Delta \Gamma = \Gamma_H - \Gamma_L  \; ,
\end{equation}
the time evolutions of the mass eigenstates are
\begin{equation}\label{time_evol_mass_states}
|D_{H, L} (t)\rangle = e^{-i(m_{H, L} - \frac{i}{2} \Gamma_{H, L})}
|D_{H, L} \rangle\; .
\end{equation}
Using Eq.~(\ref{defn_mass_eigen_state}) and Eq.~(\ref{time_evol_mass_states}), 
the time evolution of an initial $D$ or $\bar{D}$ state becomes
\begin{eqnarray} 
|D(t)\rangle &=&  g_+(t) |D \rangle  - \frac{q}{p} g_-(t) 
|\bar{D}\rangle \; , \label{timeevol-1}\\
|\bar{D}(t)\rangle &=&  g_+(t) |\bar{D} \rangle  - \frac{p}{q} g_-(t) 
|D \rangle \; , \label{timeevol-2}
\end{eqnarray}
where the coefficients $g_{\pm}(t)$ are 
\begin{eqnarray}\label{timeevolcoeff}
g_{\pm}(t) = \frac{1}{2} \left( e^{-i m_H t - \frac{1}{2} \Gamma_H t} 
\pm  e^{-i m_L t - \frac{1}{2} \Gamma_L t}\right) \; .
\end{eqnarray}
We use the standard convention
\begin{equation}
A_f = \langle f|H|D \rangle \; , \quad 
\bar{A}_f = \langle f|H|\bar{D} \rangle
\label{eq:af-abarf}
\end{equation}
to denote the amplitudes for the decay of $D$ and $\bar{D}$ mesons 
to a final state $f$.
The time-dependent decay rate of an initial $D$ meson to the
final state $f$ can then be written as
\begin{equation}
\frac{d \Gamma (D(t)\rightarrow f)}{dt} = 
N_f |\langle f|H|D (t)\rangle|^2 \; ,
\end{equation}
where $N_f$ is the time-independent normalization factor. 
Using Eqs.~(\ref{timeevol-1}), (\ref{timeevol-2}), and (\ref{timeevolcoeff}), 
we get
\begin{eqnarray}\label{decayrate1}
\frac{d\Gamma (D^0(t)\rightarrow f)}{dt}  & = & 
\frac{N_f}{2} e^{-\Gamma t} |A_f|^2 \times \nonumber \\
& & \Big[ (1 + |\lambda_f|^2) \cosh(y \Gamma t) 
+ (1 - |\lambda_f|^2) \cos(x \Gamma t) \nonumber \\
&& + 2 ~ {\rm Re}(\lambda_f) \sinh(y \Gamma t) - 
2 ~ {\rm Im}(\lambda_f) \sin(x\Gamma t)\Big] \; , 
\end{eqnarray}
where $\Gamma = (\Gamma_H+\Gamma_L)/2$, and 
$\lambda_f  = (q/p)(\bar{A}_f/A_f)$.
Similarly,
\begin{eqnarray}\label{decayrate2}
\frac{d\Gamma (\bar{D}(t) \rightarrow f)}{dt} & = &
\frac{N_f}{2} e^{-\Gamma t} \left|\frac{p}{q} A_f\right|^2 \times \nonumber \\
& & \Big[ (1 + |\lambda_f|^2) \cosh(y \Gamma t) - (1 - |\lambda_f|^2) 
\cos(x \Gamma t) \nonumber \\
& & + 2 ~ {\rm Re}(\lambda_f) \sinh(y \Gamma t) 
+ 2 ~ {\rm Im}(\lambda_f) \sin(x\Gamma t) \Big] \; .
\end{eqnarray}
The expressions above are applicable for both the final states, 
$f= \pp$ and $f = \kk$, that are the focus of this paper.

\subsection{Direct and indirect CP asymmetries}
\label{sec:acp}

The time-dependent CP asymmetry for the decay process $D \to f$ is
\begin{equation}
A_{CP}(t) = \frac{ \quad \frac{d\Gamma (D(t) \rightarrow f)}{dt} 
- \frac{d\Gamma (\bar{D}(t)\rightarrow f)}{dt} \quad}
{\frac{d\Gamma
(D(t)\rightarrow f)}{dt} + 
\frac{d\Gamma (\bar{D}(t)\rightarrow f)}{dt}} \; .
\end{equation}
From Eqs.~(\ref{decayrate1}) and (\ref{decayrate2}), we get
\begin{equation}
A_{CP}(t) = \frac{\left( \left| \frac{q}{p} \right|^2-1 \right) \Omega_+ 
+ \left( \left| \frac{q}{p} \right|^2+ 1 \right) \Omega_-}{
\left( \left| \frac{q}{p} \right|^2 + 1 \right) \Omega_+ 
+ \left( \left| \frac{q}{p} \right|^2- 1 \right) \Omega_-} \; ,
\label{acp-full-expression}
\end{equation}
where 
\barr
\Omega_+ & \equiv &  (1 + |\lambda_f|^2) \cosh(y \Gamma t)  
+ 2 ~ {\rm Re}(\lambda_f) \sinh(y \Gamma t) \; , \nonumber \\ 
\Omega_- & \equiv & (1 - |\lambda_f|^2) \cos(x \Gamma t) 
- 2 ~ {\rm Im}(\lambda_f) \sin(x\Gamma t) \; .  
\earr


Since $x, y \lesssim {\cal O}(1\%)$ \cite{hfag} and 
$\Gamma t \sim {\cal O}(1)$ or less,
the above exact expression may be simplified by expanding in the  
small parameters $x$ and $y$, and keeping the leading terms. 
For convenience, we also use the notation
\begin{equation}\label{zeta}
\left|q/p\right|^2 = 1 + \zeta \; .
\end{equation}
Given the current $95\%$ bounds  $0.44 < |q/p| < 1.07$ \cite{hfag}, 
we cannot take $\zeta$ to be a small quantity.
The expansion in small parameters $x$ and $y$ to linear order allows us
to write Eq.~(\ref{acp-full-expression}) in the form\footnote{
Note that this indeed is the definition of $\acpindir$ used in the
experimental analysis of data \cite{cdf-acp-12,lhcb-acp-13-mu}.}
\begin{equation}
\acp(t) = \acpdir + \frac{t}{\tau_D} \acpindir \; , 
\label{acp-dir-indir}
\end{equation}
where $\tau_D$ is the lifetime of the $D$ meson. 
Here $\acpdir$ and $\acpindir$ are given as
\begin{eqnarray}
\acpdir &=& \frac{1-|\lambda_f|^2+ \zeta}{1+|\lambda_f|^2+\zeta} 
= \frac{1 - |\overline{A}_f/A_f|^2}{1 + |\overline{A}_f/A_f|^2} 
= \frac{1 - |R_f|^2}{1 + |R_f|^2} 
\; , \label{cpdir} \\
\acpindir &=& 
-2 \left| \frac{q}{p} \right|^2 
\frac{\left[ (1 + |\lambda_f|^2) ~ x ~ {\rm Im}(\lambda_f) +
(1 - |\lambda_f|^2) ~ y ~ {\rm Re}(\lambda_f) \right]}
{ \left( \left| \frac{q}{p} \right| ^2 +  |\lambda_f|^2 \right)^2 }\; .
\label{cpindir}
\end{eqnarray}
The above expression for $\acpindir$ reduces to the one commonly
used \cite{hfag} in the limit $R_f=1$ 
in both the decay modes, since for CP-even final states like $\pp$ 
and $\kk$, we have $\lambda_f = - |(q/p) R_f| e^{i\phi}$ 
\cite{grossman-kagan-nir,kagan-sokoloff}. 
Our expression is more general and needs to be used if
the possibility of direct CP violation is to be taken into account.
Even if the measured direct CP violation is very small, i.e.
$|R_f| \approx 1.00$, it is possible that the phase of $R_f$ is different
for the two final states. This would make the value of $\lambda_f$ 
different for the two final states. Indeed, as will be seen in
Sec.~\ref{numerical}, the measurements indicate 
$|R_\pi| \approx |R_K| \approx 1.00$ to within $1\%$, 
while ${\rm Arg}(R_\pi) - {\rm Arg}(R_K)$ can be large. 
Such a scenario would, of course, need the NP contribution to be of 
a very specific magnitude and phase. This important issue will be 
discussed later in Sec.~\ref{implications} in detail.

Note that the effective lifetimes of the decay modes $D \to f$ and
$\bar{D} \to f$ differ from $\tau_D$ by terms of ${\cal O}(x,y)$. However
this does not change $\acpdir$, and the change in $\acpindir$ due to this 
is quadratic in $x, y$. Hence this difference can be neglected in our 
linear expansion.
Integrating Eq.~(\ref{acp-dir-indir}) over the observed normalized 
distribution of the proper decay time as measured in the $D \to f$ decay,
we get
\begin{eqnarray}
\langle \acp \rangle
&=& \acpdir  + \frac{\langle t \rangle}{\tau_D}  \acpindir \; .
\end{eqnarray}
Here $\langle t \rangle$ is average decay time that can be measured separately
for each $D \to f$ decay mode. For the $\pp$ and $\kk$ decay modes,
\begin{eqnarray}
\langle \acp(\pi) \rangle &=& \acpdir(\pi) + \frac{\langle t(\pi) \rangle}{\tau_D} 
\acpindir(\pi) \; , \\
\langle \acp(K) \rangle &=& \acpdir(K) + \frac{\langle t(K) \rangle}{\tau_D} 
\acpindir(K) \; , 
 \end{eqnarray}
where $\langle t(\pi) \rangle,\langle t(K)\rangle$ are average decay times 
of $D$ mesons for decays into the $\pp$ and $\kk$ states, respectively.
These average times are characteristics of specific experiments,
the values for which have been shown in Table~\ref{tab:times}.
Note that for the LHCb data,
$\langle \bar{t} \rangle = (\langle t(K) \rangle +
\langle t(\pi) \rangle)/2 $ and  
$\Delta \langle t \rangle = \langle t(K) \rangle - 
\langle t(\pi) \rangle$.

\begin{table}
\centering
\begin{math}
\begin{array}{ccr}
\hline
\text{Quantity} & \text{Value} & \text{Reference} \\
\hline
\langle t (\pi) \rangle/\tau_D
& 2.4 \pm 0.03 & \text{CDF 2011*, CDF 2012} ~\cite{cdf-acp-11,cdf-acp-12}\\
\langle t (K) \rangle/\tau_D
& 2.65 \pm 0.03 & \text{CDF 2011*, CDF 2012} ~\cite{cdf-acp-11,cdf-acp-12}\\
\hline
\Delta \langle t \rangle/\tau_D
& 0.0983 \pm 0.0022 \pm 0.0019 & \text{LHCb 2011}* (\pi \text{-tagged}) 
 ~\cite{lhcb-acp-11} \\
& 0.1119 \pm 0.0013 \pm 0.0017 & \text{LHCb 2013} (\pi \text{-tagged}) 
~\cite{lhcb-acp-13-pi} \\
& 0.018 \pm 0.002 \pm 0.007 & \text{LHCb 2013} (\mu \text{-tagged}) 
~\cite{lhcb-acp-13-mu} \\
\hline
\langle \bar{t} \rangle / \tau_D
& 2.0826 \pm 0.0077 & \text{LHCb 2011}* (\pi \text{-tagged}) 
~\cite{lhcb-acp-11}\\
& 2.1048 \pm 0.0077 & \text{LHCb 2013} (\pi \text{-tagged}) 
~\cite{lhcb-acp-13-pi}\\
& 1.062 \pm 0.001 \pm 0.003 & \text{LHCb 2013} (\mu \text{-tagged}) 
~\cite{lhcb-acp-13-mu}\\
\hline
\end{array}
\end{math}
\caption{Experimental values of various quantities that appear in the
determination of $\acpdir$ and $\acpindir$. We take
$\tau_D=(0.41 \pm 0.0015)$ ps \cite{pdg-2012}. 
The data marked with a * are not used for the fit.} 
\label{tab:times}
\end{table}

Subsequently, the  difference CP asymmetry can be obtained as
\cite{gersabeck-12}
\begin{eqnarray}
\label{delta-acp}
\Delta \acp &=& \langle \acp(K) \rangle  
- \langle \acp (\pi) \rangle  \\ \nonumber
&=& \acpdir (K) - \acpdir (\pi)  +
\frac{\langle t(K) \rangle}{\tau_D} \acpindir(K) 
- \frac{\langle t(\pi) \rangle}{\tau_D} \acpindir(\pi) \; .
\end{eqnarray}
As can be seen from Eqs.~(\ref{cpdir}) and (\ref{cpindir}), 
$\acpindir$ depends on the final state $f$ through its $\lambda_f$. 
Hence in general, indirect CP asymmetries in $D \to \pp$ 
and $D \to \kk$ can be different.
Hence the equality $\acpindir(\pi)=\acpindir(K)$, as is generally 
used in the analyses of these channels, is only approximate. 
Note that this statement is independent of the mechanism of CP violation,
since our analysis has been completely model-independent.

We would like to make a subtle point here. Equation~(\ref{cpindir}) indeed
agrees with the statement made in the literature \cite{grossman-kagan-nir}
that in the absence of direct CP violation, the indirect CP violation is 
universal. However this statement needs to be interpreted with caution. 
It is true only if the absence of direct CP violation is taken to mean 
$R_f=1$, both in magnitude as well as phase, for all modes (in any consistent 
phase convention). The absence of observable direct CP violation, however, 
only requires $|R_f|=1$, which is not enough to guarantee this universality.
On the other hand, the universality of indirect CP violation only needs 
$R_f$ to be equal (in magnitude as well as phase) for all relevant decay 
modes, and its magnitude is immaterial.

\begin{table}
\centering
\begin{math}
\begin{array}{clr}
\hline
\text{Quantity} & \text{Value (\%)} & \text{Reference} \\
\hline
\langle \acp(\pi) \rangle & 
-4.9 \pm 7.8 \pm 3.0 & \text{E791 1997*} ~\cite{e791-acp-97} \\ 
& 4.8 \pm 3.9 \pm 2.5 & \text{FOCUS 2000*} ~\cite{focus-acp-00} \\
& 1.9 \pm 3.2 \pm 0.8 & \text{CLEO 2001*} ~\cite{cleo-acp-01} \\
& 0.04 \pm 0.69 & \text{CDF 2011*} ~\cite{cdf-acp-11}\\
& -0.24 \pm 0.52 \pm 0.22 & \text{Babar 2008} 
~\cite{babar-acp-08} \\
& 0.55 \pm 0.36 \pm 0.09 & \text{Belle 2012} ~\cite{belle-acp-12} \\
\hline
\langle \acp(K) \rangle 
&  - 1.0 \pm 4.9 \pm 1.2 & \text{E791 1997*} ~\cite{e791-acp-97} \\ 
& -0.1 \pm 2.2 \pm 1.5 & \text{FOCUS 2000*} ~\cite{focus-acp-00} \\
& 0.0 \pm 2.2 \pm 0.8 & \text{CLEO 2001*} ~\cite{cleo-acp-01} \\
& 0.00 \pm 0.34 \pm 0.13 & \text{Babar 2008}  ~\cite{babar-acp-08} \\
& -0.24 \pm 0.41 & \text{CDF 2011*} ~\cite{cdf-acp-11}\\
&  -0.32 \pm 0.21 \pm 0.09 & \text{Belle 2012} ~\cite{belle-acp-12} \\  
\hline
\Delta \acp 
& -0.82 \pm 0.21 \pm 0.11 & \text{LHCb 2011*} ~ (\pi \text{-tagged}) 
~\cite{lhcb-acp-11} \\
& -0.62 \pm 0.21 \pm 0.10 & \text{CDF 2012} ~(\pi \text{-tagged})
~\cite{cdf-acp-12}\\
& -0.34 \pm 0.15 \pm 0.10 & \text{LHCb 2013}~ (\pi \text{-tagged})
~\cite{lhcb-acp-13-pi} \\ 
& 0.49 \pm 0.30 \pm 0.14 & \text{LHCb 2013}~ (\mu \text{-tagged})
~\cite{lhcb-acp-13-mu} \\
\hline
\end{array}
\end{math}
\caption{Experimental values of CP asymmetries measured at the experiments.
The data marked with a * are not used for the fit.} 
\label{tab:acp}
\end{table}

The measured values of $\langle \acp(\pi) \rangle$, 
$\langle \acp(K) \rangle$, and $\Delta \acp$ are shown in 
Table~\ref{tab:acp}. 
The assumption of $\acpindir(\pi)= \acpindir(K)$ may
also be responsible for the apparent discrepancy between the values of
$\Delta \acp$ measured at the LHCb through the pion-tagged and the
muon-tagged samples. 
Note that the values of
$\langle t(\pi) \rangle = \langle \bar{t} \rangle - \Delta \langle t
\rangle/2$ for the two samples are different, and so are the values of
$\langle t(K) \rangle = \langle \bar{t} \rangle + \Delta \langle t 
\rangle/2$. Indeed, we can write the difference
$\delta (\Delta \acp) \equiv (\Delta {\acp})_{\pi} - (\Delta {\acp})_{\mu}$
as
\barr
\delta(\Delta\acp) & = &  
\left( \frac{ \langle t(K) \rangle_{\pi} - \langle t(K) \rangle_{\mu}}
{\tau_D} \right) \acpindir(K) -
\left( \frac{ \langle t(\pi) \rangle_{\pi} - \langle t(\pi) \rangle_{\mu}}
{\tau_D} \right) \acpindir(\pi)   \nonumber \\
& = &  \left( \frac{ (\Delta \langle t \rangle)_\pi -
(\Delta \langle t \rangle)_\mu }{2 \tau_D} \right)
 \left[ \acpindir(K) + \acpindir(\pi) \right] + \nonumber \\
&  &  \hspace{2cm} \left( \frac{ \langle \bar{t} \rangle_\pi -
\langle \bar{t} \rangle_\mu }{\tau_D} \right)
\left[ \acpindir(K) - \acpindir(\pi) \right] \; . 
\label{eq:delta-delta-acp}
\earr
The term on the last line would be missed if one assumes 
$\acpindir(\pi)= \acpindir(K)$. We shall revisit this
quantitatively during our numerical analysis in the next section.

\subsection{CP violating observables through effective lifetimes}
\label{sec:ycp}

The expansion of Eq.~(\ref{decayrate1}) to first order in 
$x, y$ yields
\begin{equation}
\frac{d\Gamma}{dt}(D(t)\rightarrow f) \approx \frac{N_f}{2} e^{-\Gamma t} 
|A_f|^2 \Big[ 1 + y \Gamma t ~{\rm Re}(\lambda_f) 
- x \Gamma t ~{\rm Im}(\lambda_f) \Big] \; . 
\end{equation}
With $z_f \equiv x ~ {\rm Im}(\lambda_f)- y ~ {\rm Re}(\lambda_f) $,
this could be written in the form \cite{Bergmann:2000id}:
\begin{equation}
\frac{d\Gamma}{dt}(D(t)\rightarrow f) \propto e^{-\Gamma t} 
(1 - z_f \Gamma t) \; .
\end{equation}
The effective lifetime in the $D \to f$ mode is then 
\begin{equation}
\tau_f \approx (1 - z_f)/\Gamma  \; . 
\end{equation}
Since $z_f$ depends on the decay mode $D \to f$ in general, the effective
lifetimes measured in different modes can be different. These differences
may be used to construct observables that are sensitive to CP violation
in $D$ decays. 

For the decay $\bar{D} \to f$, Eq.~(\ref{decayrate2}) may 
also be written in another convenient form
\begin{eqnarray}\label{decayrate2-2}
\frac{d\Gamma}{dt}(\bar{D}(t) \rightarrow f) & = &
\frac{N_f}{2} e^{-\Gamma t} \left|\bar{A}_f\right|^2 \times \nonumber \\
& & \Big[ (1 + |\lambda_f^{-1}|^2) \cosh(y \Gamma t) + (1 - |\lambda_f^{-1}|^2) 
\cos(x \Gamma t) \nonumber \\
& & + 2 ~ {\rm Re}(\lambda_f^{-1}) \sinh(y \Gamma t) 
- 2 ~ {\rm Im}(\lambda_f^{-1}) \sin(x\Gamma t) \Big] \; .
\end{eqnarray}
After neglecting terms that are quadratic or higher powers in $x,y$,
one gets
\begin{equation}
\frac{d\Gamma}{dt}(\bar{D}(t)\rightarrow f) \propto e^{-\Gamma t} 
(1 - \bar{z}_f \Gamma t ) \; ,
\end{equation}
with $\bar{z}_f \equiv x {\rm Im}(\lambda_f^{-1}) -y {\rm Re}(\lambda_f^{-1})$,
so that the effective lifetime for this mode becomes
\begin{equation}
\bar{\tau}_f = (1 - \bar{z}_f)/\Gamma  \; . 
\end{equation}

When the final state $f$ is a CP eigenstate $\fcp$ like $\pp$ or $\kk$, 
a CP-violating quantity can be constructed from the difference 
of the effective lifetimes of $D \to \fcp$ and  $\bar{D} \to \fcp$
\cite{Bergmann:2000id}:
\barr
\agamma(\fcp) & \equiv & \frac{\bar{\tau}_\fcp - \tau_\fcp}{
\bar{\tau}_\fcp + \tau_\fcp}  
= \frac{1}{2} (z_\fcp - \bar{z}_\fcp) \nonumber \\
& = &  \frac{1}{2} \bigg (
x [{\rm Im}(\lambda_\fcp) - {\rm Im}(\lambda_\fcp^{-1})] 
-y [{\rm Re}(\lambda_\fcp) - {\rm Re}(\lambda_\fcp^{-1})] \bigg) \; 
\label{eq:agamma}
\earr
clearly vanishes in the limit of CP conservation since $\lambda_\fcp =\pm 1$
in that case. This expression reduces to the one used in Ref.~\cite{hfag} in 
the limit $R_f=1$, as expected.
The relation $\agamma = -\acpindir$ used in Ref.~\cite{hfag} is also valid
only in this approximation, the actual relation being
\beq
\agamma = - \frac{1}{4} \acpindir \left( |R_f| + \frac{1}{|R_f|} \right)^2 \;. 
\eeq
Although the form of Eq.~(\ref{eq:agamma}) seems different from the 
one given in Ref.~\cite{gersabeck-12}, it is a result of expansions up to
different orders in small quantities. In particular, we do not
assume $\zeta (\equiv |q/p|^2-1)$ to be small, and keep terms to a higher
power in it. 

The quantities $\agamma(\pi)$ and $\agamma(K)$ have been measured
separately \cite{belle-ycp-07,babar-ycp-08,lhcb-ycp-11} and as an
average over the two modes \cite{belle-ycp-12,babar-ycp-12};
however, the errors are
not small enough for a nonzero measurement. See Table~\ref{tab:ycp}.

\begin{table}
\centering
\begin{math}
\begin{array}{clr}
\hline
\text{Quantity} & \text{Value (\%)} & \text{Reference} \\
\hline
A_{\Gamma}(\pi) 
& -0.28 \pm 0.52 \pm 0.15 & \text{Belle 2007} ~\cite{belle-ycp-07} \\
& -0.049 \pm 0.73 & \text{Babar 2008} ~\cite{babar-ycp-08} \\
\hline
A_{\Gamma}(K) 
& 0.15 \pm 0.35 \pm 0.15 & \text{Belle 2007} ~\cite{belle-ycp-07} \\
& 0.39 \pm 0.46 & \text{Babar 2008} ~\cite{babar-ycp-08} \\
& -0.59 \pm 0.59  \pm 0.21 & \text{LHCb 2011} ~\cite{lhcb-ycp-11}\\
\hline
A_{\Gamma}({\rm avg})
& -0.03 \pm 0.20 \pm 0.08 & \text{Belle 2012} ~\cite{belle-ycp-12} \\
& 0.09 \pm 0.27 & \text{Babar 2013} ~\cite{babar-ycp-12} \\
\hline
\ycp(\pi) 
& 0.5 \pm 4.3 \pm 1.8 & \text{CLEO 2001} ~\cite{cleo-acp-01} \\
& 1.44 \pm 0.57 \pm 0.25 & \text{Belle 2007} ~ \cite{belle-ycp-07} \\
& 0.46 \pm 0.65 \pm 0.25 & \text{Babar 2008} ~\cite{babar-ycp-08} \\
\hline
\ycp(K) 
& 3.42 \pm 1.39 \pm 0.74 & \text{Focus 2000} ~\cite{focus-ycp-00} \\
& -1.9 \pm 2.9 \pm 1.6 & \text{CLEO 2001} ~\cite{cleo-acp-01} \\
& 1.25 \pm 0.39 \pm 0.25 & \text{Belle 2007} ~\cite{belle-ycp-07} \\
& 1.60 \pm 0.46 \pm 0.17 & \text{Babar 2008} ~\cite{babar-ycp-08} \\
& 1.12 \pm 0.26 \pm 0.22 & \text{Babar 2009} ~\cite{babar-ycp-09} \\
& 0.55 \pm 0.63 \pm 0.41 & \text{LHCb 2011} ~\cite{lhcb-ycp-11}\\
\hline
\ycp({\rm avg}) 
& 1.11 \pm 0.22 \pm 0.11 & \text{Belle 2012} ~\cite{belle-ycp-12} \\
& 0.72 \pm 0.18 \pm 0.12 & \text{Babar 2013} ~\cite{babar-ycp-12} \\
\hline
\end{array}
\end{math}
\caption{Measured values of $\agamma$ and $\ycp$.}
\label{tab:ycp}
\end{table}

For flavor-specific decays, where $D \to f$ is allowed
but $\bar{D} \to f$ is not, $\lambda_f$ vanishes and Eq.~(\ref{decayrate1})
gives
\beq
\frac{d\Gamma}{dt}(D(t) \to f) \approx
N_f e^{-\Gamma t} |A_f|^2 \; ,
\eeq
when terms with quadratic and higher powers of $x, y$ are
neglected. The average lifetime for such processes is clearly
$\tau_{\rm FS} \approx 1/\Gamma$.
(Note that while taking $D \to \pi^- K^+$ to be a flavor-specific mode,
the doubly Cabibbo-suppressed decay $D \to \pi^+ K^-$ has been neglected.)
The quantity 
\barr
\ycp & \equiv & \frac{\tau_{FS}}{(\tau_{\fcp}+ \bar{\tau}_{\fcp})/2} -1
\approx \frac{1}{2} \left( z_\fcp + \bar{z}_\fcp \right) \nonumber \\
& = &  \frac{1}{2} \bigg (
x [{\rm Im}(\lambda_\fcp) + {\rm Im}(\lambda_\fcp^{-1})] 
-y [{\rm Re}(\lambda_\fcp) + {\rm Re}(\lambda_\fcp^{-1})] \bigg) \; 
\label{eq:ycp}
\earr
is not necessarily CP-violating; however, it forms an important input 
for disentangling  $z_\fcp$ and $\bar{z}_\fcp$ from the measurement of 
$\agamma$. Note that the expression clearly reduces to the
one used in Ref.~\cite{hfag} in the limit of $\bar{A}_f=A_f$ (or $R_f=1$), and
in the absence of CP violation, i.e. $\phi=0$, one has $y_{\rm CP}=y$.

Taking the CP-eigenstates to be $\pp$ and the flavor-specific
final state to be $\pi K$, The measurements of $\ycp(\pi)$ have so far 
been consistent with zero to within $2\sigma$
\cite{cleo-acp-01,belle-ycp-07,babar-ycp-08}, while 
taking the CP-eigenstates to be $\kk$, nonzero measurements of
$\ycp(K)$ to more than $4\sigma$ level have been obtained 
\cite{belle-ycp-07,babar-ycp-08,babar-ycp-09}.
Averaging over $\pp$ and $\kk$ modes \cite{belle-ycp-12,babar-ycp-12}
also gives a nonzero value at more than the $3\sigma$ level.

\section{Numerical analysis}
\label{numerical}

We now perform a $\chi^2$ fit to the data on the $\ddbar$
mixing and decay with an aim toward disentangling the contributions
from CP-violation in mixing and in decay.
The fit is performed to the four model-independent complex parameters
$M_{12}, \Gamma_{12}, R_\pi$ and $R_K$.
Here one has to be careful about the data to be included. We use the
following prescription:

\begin{itemize}

\item For the data on $\langle \acp(\pi) \rangle, 
\langle \acp(K) \rangle$ and $\Delta \acp$,
we use the experimental data as shown in Table~\ref{tab:acp} 
directly. The average decay times as given in Table~\ref{tab:times} are used.
Note that the data marked with a * are shown for the sake of
completeness, but they are not used in the fit, either because they
give too weak constraints, or because they have been used in later results
by the same collaboration.
This helps avoid double counting the same data.

\item For the data on $\agamma$ and $\ycp$, we do not use the
COMBOS fit \cite{hfag} directly since it assumes equal values
of these quantities in the $\pp$ and $\kk$ channels.
Whereas, as can be seen from Sec.~\ref{sec:ycp}, the difference 
between $\agamma(\pi)$ and $\agamma(K)$, as well as between 
$\ycp(\pi)$ and $\ycp(K)$, is of linear order in $x,y$ when 
$\lambda_{\pi} \neq \lambda_K$, and hence is relevant for our analysis here.
We therefore use the data on $\agamma(\pi), \agamma(K), \ycp(\pi)$
and $\ycp(K)$ separately, as shown in Table~\ref{tab:ycp}.
We have to contend with the problem that the most recent Belle
and Babar results \cite{belle-ycp-12,babar-ycp-12} 
only report values of $\agamma$ and $\ycp$ that are averaged over
the $\pp$ and $\kk$ modes. 
While using these data, we take the averaged values to be 
weighted averages, with weights proportional to the number of events
in the two modes. However it would have been desirable to
have the values of $\agamma(\pi), \agamma(K), \ycp(\pi)$ and
$\ycp(K)$ separately, directly from the experiments, for a more 
accurate analysis.

\begin{table}
\centering
\begin{math}
\begin{array}{clr}
\hline
\text{Quantity} & \text{Value} & \text{Reference} \\
\hline
x & (0.80 \pm 0.29 ^{+0.09 + 0.10} _{-0.07-0.14})\% & \text{Belle 2007} 
~\cite{belle-ddbar-07} \\
y & (0.33 \pm 0.24 ^{+0.08 + 0.06} _{-0.12 -0.08})\%   & \text{Belle 2007} 
~\cite{belle-ddbar-07} \\ 
|q/p| & 0.86 ^{+0.30 + 0.06}_{-0.29 - 0.03} & \text{Belle 2007} 
~\cite{belle-ddbar-07} \\
\varphi & (-14 ^{+16 + 5 + 2}_{-18-3-4}) \text{ degrees} & \text{Belle 2007} 
~\cite{belle-ddbar-07} \\
\hline
x & (0.16 \pm 0.23 \pm 0.12 \pm 0.08)\% & \text{Babar 2010} 
~\cite{babar-ddbar-10} \\
y & (0.57 \pm 0.20 \pm 0.13 \pm 0.07)\% & \text{Babar 2010} 
~\cite{babar-ddbar-10} \\
\hline
R_M = (x^2 + y^2)/2 & 0.0130 \pm 0.0269 & \cite{hfag} \\
\hline
\end{array}
\end{math}
\caption{Experimental input for $\ddbar$ mixing parameters.}
\label{tab:ddbar}
\end{table}

\item For the input from $\ddbar$ mixing also, we do not use the 
HFAG \cite{hfag} fit for $x,y, |q/p|, \varphi$ directly here since 
in addition to the $\ddbar$ mixing data, 
it uses the data on $\acp, \agamma$ and $\ycp $ that have
already been used above, which would have given rise to
double counting of the data. 
We therefore only use the data from \cite{belle-ddbar-07,babar-ddbar-10}, 
and the COMBOS average \cite{hfag} for $R_M \equiv (x^2+y^2)/2$ 
from the semileptonic $D$ decays.
To keep the number of fit parameters limited, we do not
use the data on mixing parameters from the $K\pi$ or $K\pi\pi$ channel.

\end{itemize}

The $\chi^2$ function is taken to be 
\begin{eqnarray}
\chi^2 &=& \sum_{i=1}^{32} \frac{(X_i - X_i^{exp})^2}{(\sigma_{X_i})^2} \; .
\end{eqnarray}
Here $X_i, X_i^{exp}$, and $\sigma_{X_i}$ with $(i=1,2,3,\dots, 32)$ 
represent the theoretical values, experimental values and corresponding 
experimental uncertainties, respectively, of the observables given
in Tables~\ref{tab:acp}, \ref{tab:ycp} and \ref{tab:ddbar}.
We add the statistical and systematic errors in quadrature,
and take all the measurements to be independent and uncorrelated.
The MINUIT \cite{James:1975dr} subroutine is used for the
minimization of $\chi^2$ in the multidimensional parameter space.
The best-fit values of the fit parameters are:
\begin{center}
\begin{tabular}{llll}
$|M_{12}| = 0.0059$ ps$^{-1}$ , &
${\rm Arg}(M_{12}) = 3.37$ ,&
$|R_\pi| = 1.002$ , &
${\rm Arg}(R_\pi) = 3.82$ , \\
$|\Gamma_{12}|= 0.0207$ ps$^{-1}$ , &
${\rm Arg}(\Gamma_{12})= 3.39$ , & 
$|R_K| = 1.000$ ,&
${\rm Arg}(R_K) = 3.20$ . \\
\end{tabular}
\end{center}
Note that the phases of $M_{12}$ and $\Gamma_{12}$ are 
convention dependent, however, the difference between them is 
independent of phase conventions. 
Since this difference is small, the magnitude of $q/p$ at the
best-fit point is still close to unity.
It is also observed that the values of best fit for
$|R_K|$ as well as $|R_\pi|$ do not deviate much from unity, so that the
direct CP violation in both these decay modes is expected to be rather small.
However, the phases of $R_\pi$ and $R_K$ at the best-fit point are 
significantly different. This will be relevant in our discussion later.

The fit is rather good: at the best-fit point, 
$\chi^2/{\rm dof} = 27.6/24$. 
It is interesting that even if we impose a further restriction of 
$|\lambda_{\pi}|= |\lambda_{K}|$, which would correspond to
$|R_\pi| =|R_K|$, 
the fit still stays almost as good, with $\chi^2/{\rm dof} = 28.1/25$.
The values of the best-fit parameters are also very similar.
This indicates that around the best-fit point, the CP-violation
through mixing alone as well as CP-violation through decay alone
is very small, so the CP violation observed is mainly through the 
interference between mixing and decay. 
Further insisting on identical values for the magnitudes as well
as phases of $\lambda_{\pi}$ and $\lambda_{K}$, however,
worsens the fit to $\chi^2/{\rm dof} = 37.3/26$.
This indicates the tension of the data with the scenario of equal CP
violation in the two decays. As indicated in the previous section,
this implies significantly different values of $\acpindir$ in the 
two decays.

To make quantitative statements about the significance of
our observations above, we show the parameter spaces favored by
the data in Fig.~\ref{fig:param}. The contours shown in the figure 
correspond to $\Delta \chi^2 = 2.3$ (68\% C.L.) and 
$\Delta \chi^2 = 4.61$ (90\% C.L.). The following observations may be 
made from the figure:

\begin{figure}[t]

\includegraphics[width=0.4\textwidth,height=0.49\textwidth,angle=-90]
{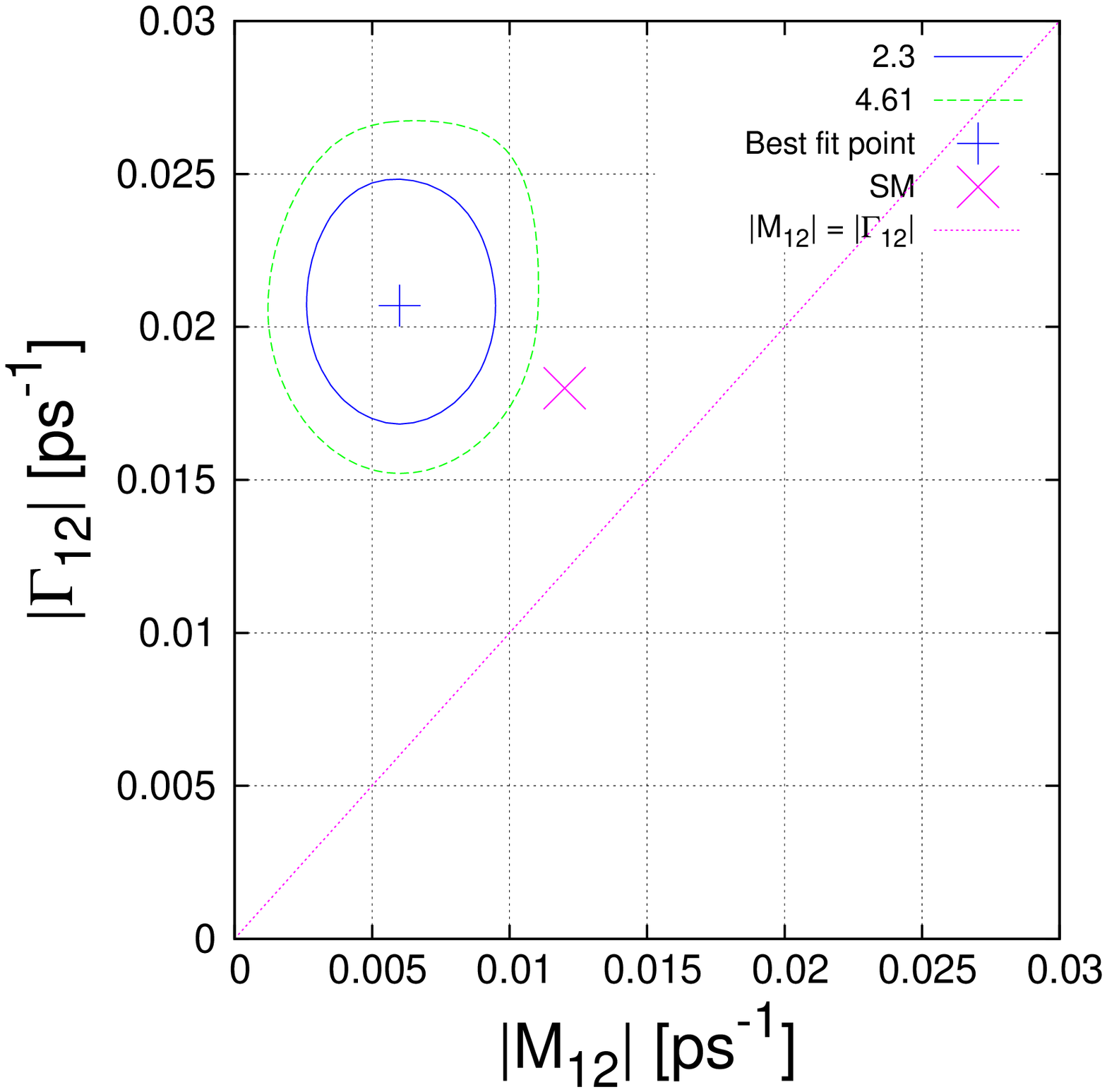}
\includegraphics[width=0.4\textwidth,height=0.49\textwidth,angle=-90]
{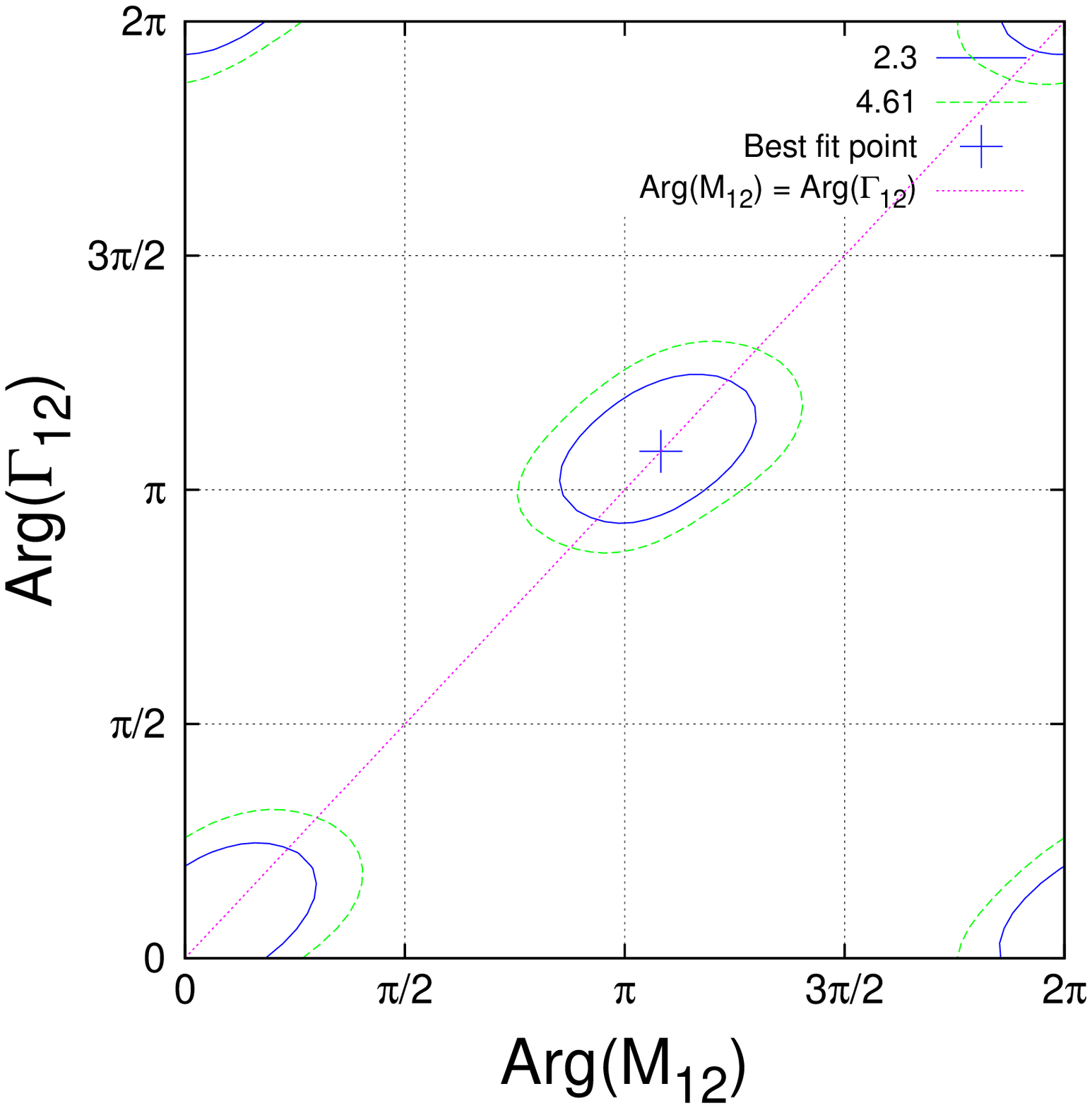}

\vspace{0.2cm}

\includegraphics[width=0.4\textwidth,height=0.49\textwidth,angle=-90]
{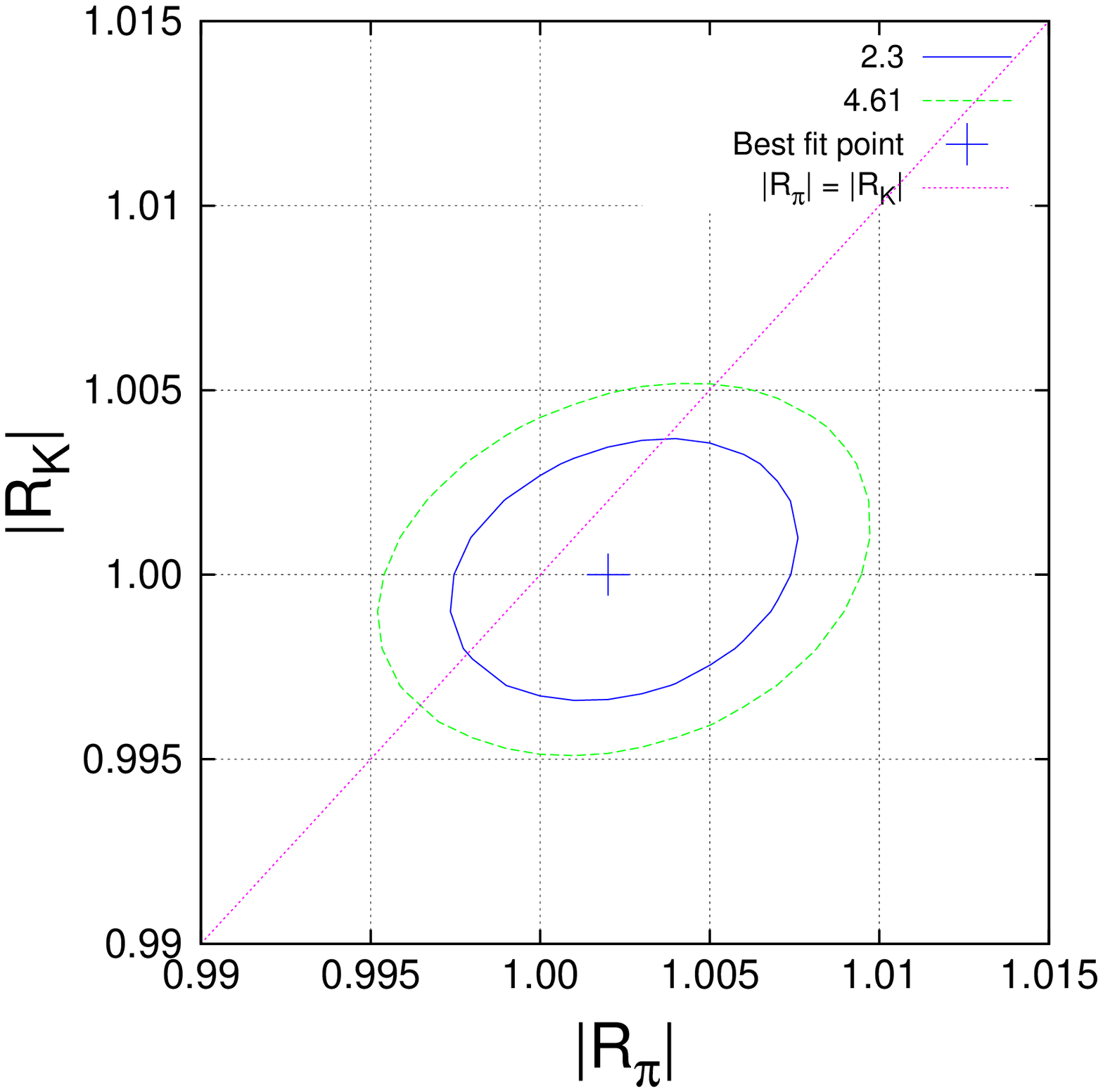}
\includegraphics[width=0.4\textwidth,height=0.49\textwidth,angle=-90]
{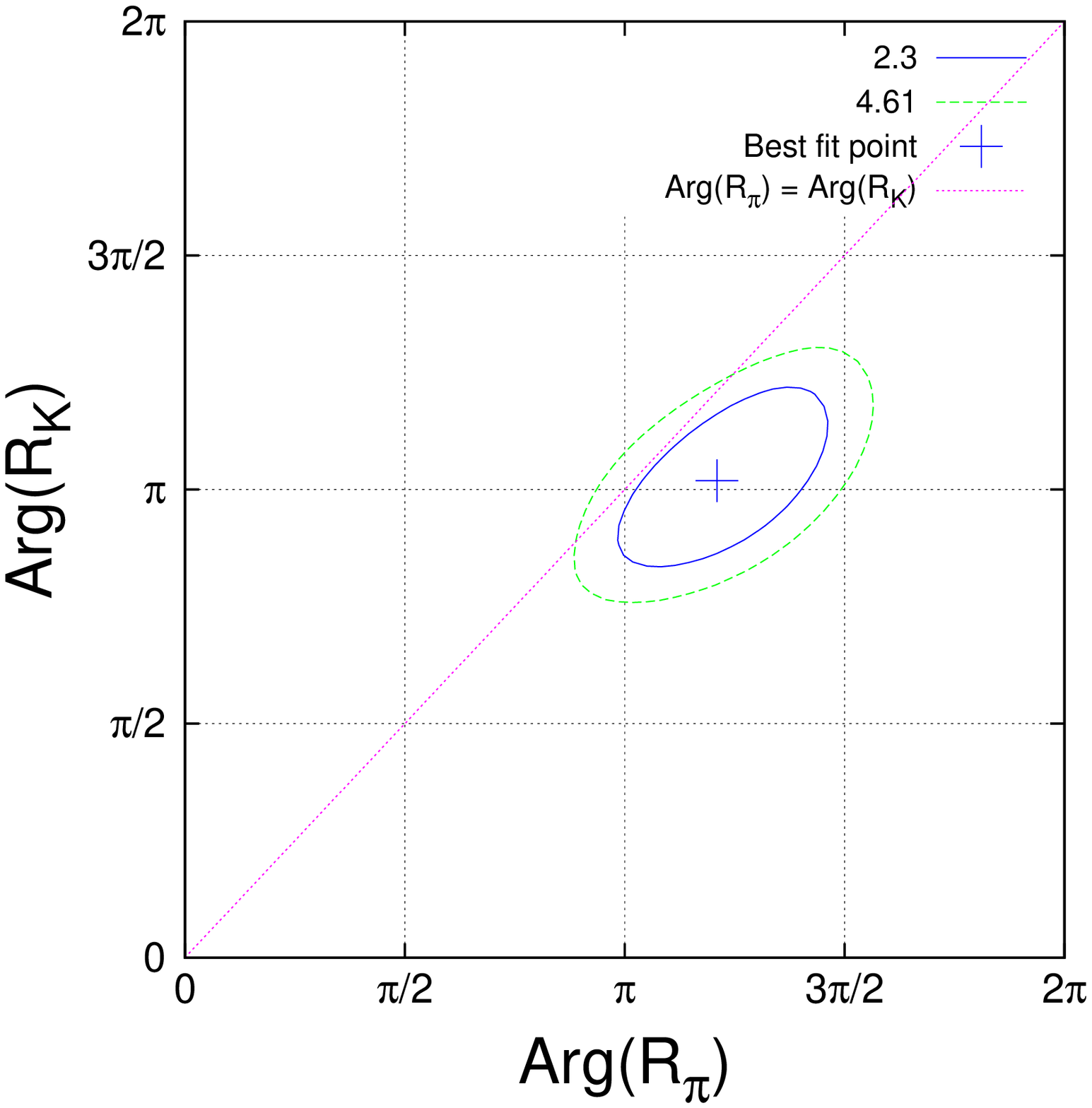}

\caption{Parameter spaces favored by the current data on $\ddbar$ mixing
and decay. While displaying constraints on two parameters, the values of
the other parameters are varied over all their allowed ranges to minimize
$\chi^2$. The SM value of $(|M_{12}|,|\Gamma_{12}|)$ is taken from 
Ref.~\cite{bigi-blanke}. 
}
\label{fig:param}
\end{figure}

\begin{itemize}
\item $|\Gamma_{12}| > |M_{12}|$ in the whole of the region 
allowed to 90\% C.L.. Indeed, the data seem to favor the region
with $|\Gamma_{12}|$ equal to a few times $|M_{12}|$. 

\item The phase between $M_{12}$ and $\Gamma_{12}$ is compatible
with zero, although deviations of up to $\approx 0.9$ radians are possible.
The SM prediction for this phase would be ${\cal O}(0.1\%)$,
so much larger values for this phase are still allowed by the data.

\item The 90\% allowed values for both $|R_\pi|$ and $|R_K|$,
for both $\pp$ and $\kk$ modes are consistent with unity.
While a deviation of $\approx 1\%$ from unity is allowed for $|R_{\pi}|$,
the value of $|R_K|$ is restricted to be within $\approx 0.5\%$ of unity.
This indicates that the direct CP violation
$\acpdir$ is restricted to a fraction of a per cent.

\item The relative phase between $R_\pi$ and $R_K$ is a 
convention-independent, physical quantity. Data seem to prefer
different phases for $R_\pi$ and $R_K$.
This indicates that although $|R_\pi|\approx |R_K|$ is
allowed, $R_\pi = R_K$ is disfavored. This would further imply
$\lambda_\pi \neq \lambda_K$, and consequently
$\acpindir(\pi) \neq \acpindir(K)$.

\end{itemize}

\begin{figure}
\begin{center}
\includegraphics[width=0.4\textwidth,height=0.49\textwidth,angle=-90]
{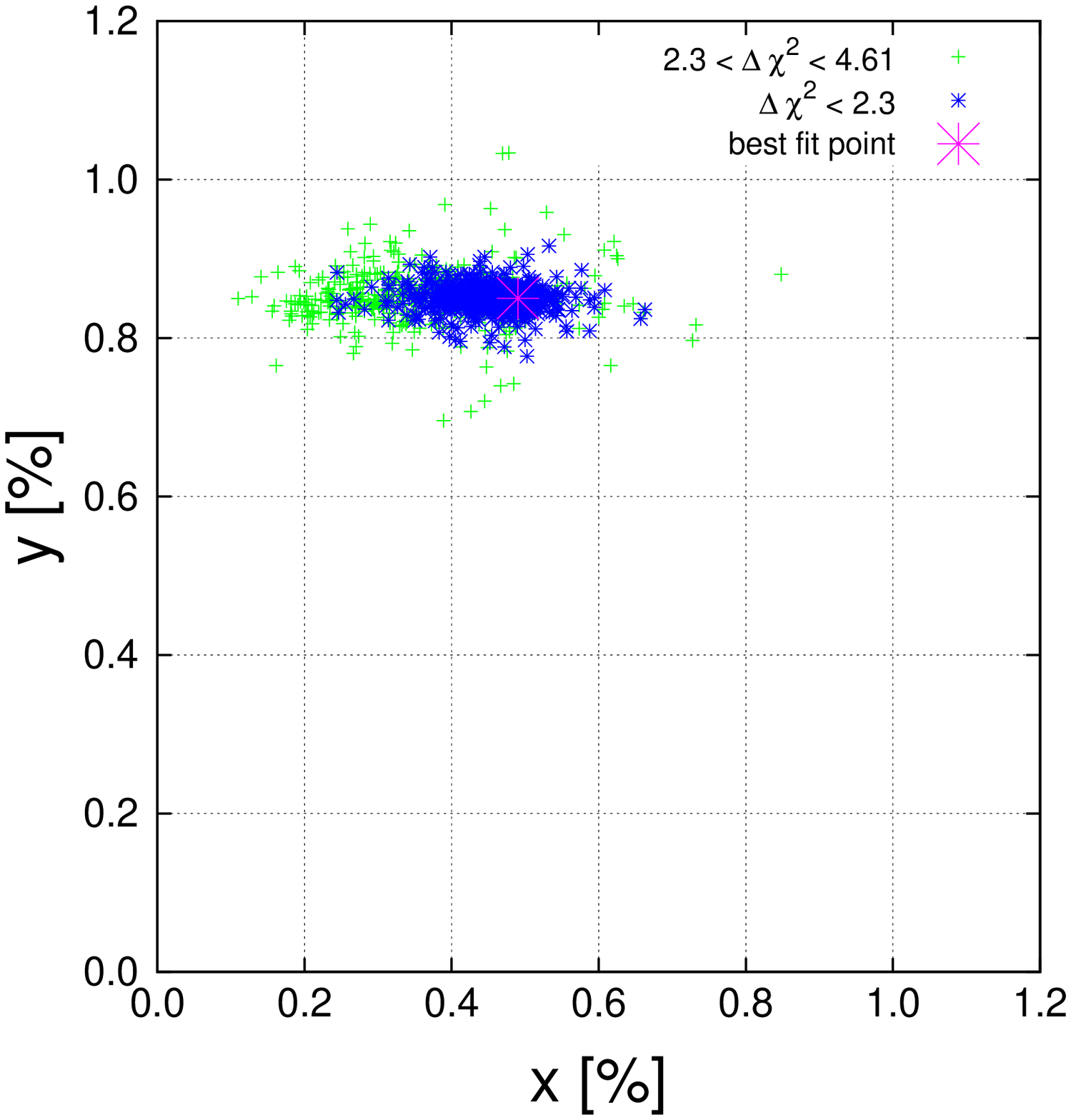}
\includegraphics[width=0.4\textwidth,height=0.49\textwidth,angle=-90]
{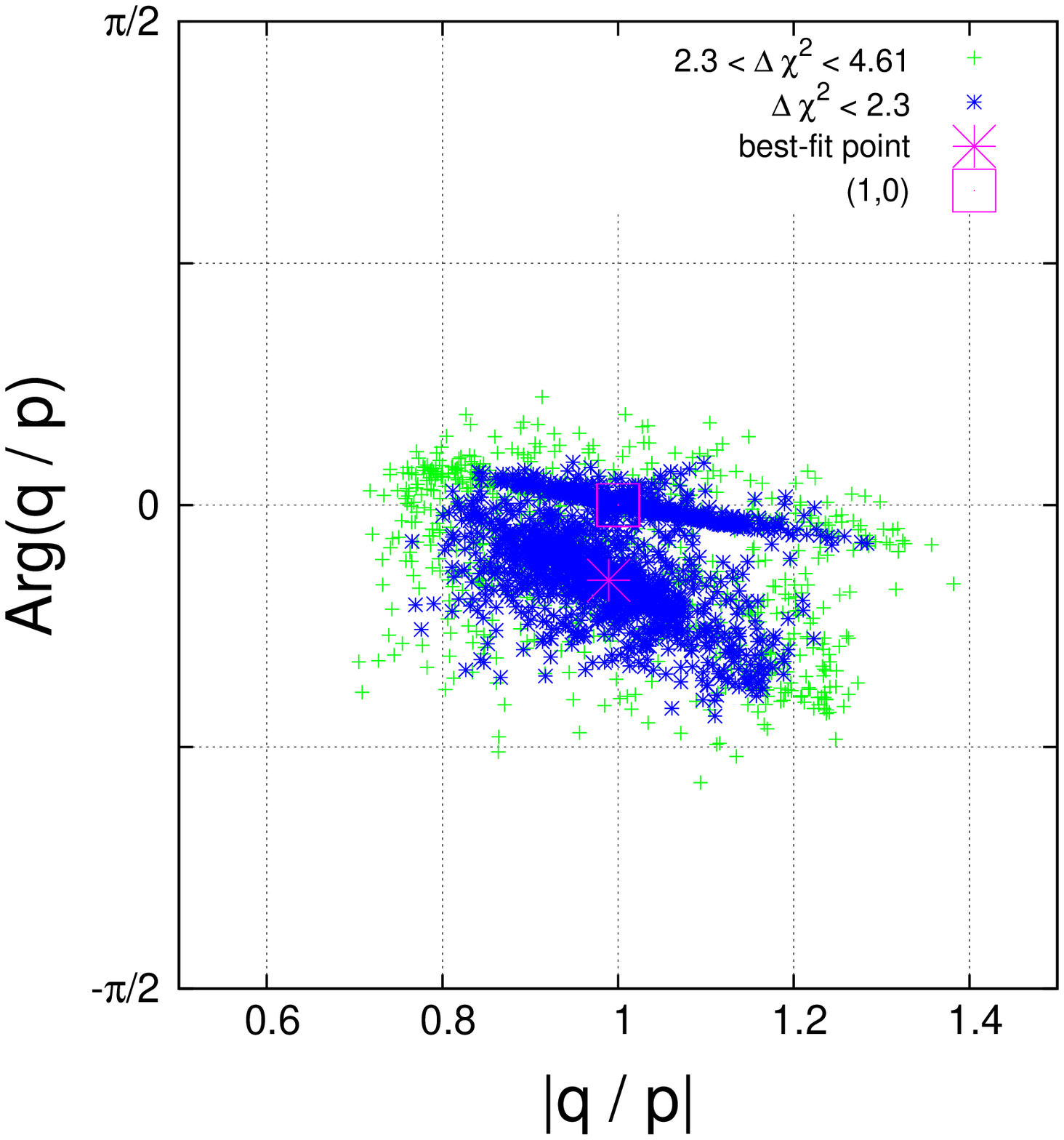}
\caption{Mixing parameters in $\ddbar$ mixing, applicable to both,
$\pp$ and $\kk$ decays.}
\label{fig:qbyp}
\end{center}
\end{figure}

The derived quantities $x,y,|q/p|$ and $\varphi\equiv {\rm Arg}(q/p)$
that are commonly used to describe the $\ddbar$ mixing are shown
in Fig.~\ref{fig:qbyp}. The figure indicates the following:
\begin{itemize}
\item The values of both $x$ and $y$ are positive to 90\% C.L..

\item Although $|q/p|$ is consistent with unity, a variation in the
range $(0.7,1.3)$  is still allowed to 90\% C.L.. 
This implies that significant CP violation through mixing is allowed.
Note that our fit gives higher values of $|q/p|$ as compared
to the one in Ref.~\cite{hfag}; however, there are differences in the two
fit procedures. We have used only a subset of the data used therein, 
but have taken care of possibly different values of $\agamma$ and $\ycp$
in the $\pp$ and $\kk$ modes.

\item The phase $\varphi$ is restricted to be 
in the range $(-0.9, 0.3)$ to 90\% C.L.. 
This quantity is of course phase-convention dependent, and is
physically meaningful only when compared with the phases of
$R_\pi$ or $R_K$. So we shall not discuss it further.

\end{itemize}

\begin{figure}

\includegraphics[width=0.4\textwidth,height=0.49\textwidth,angle=-90]
{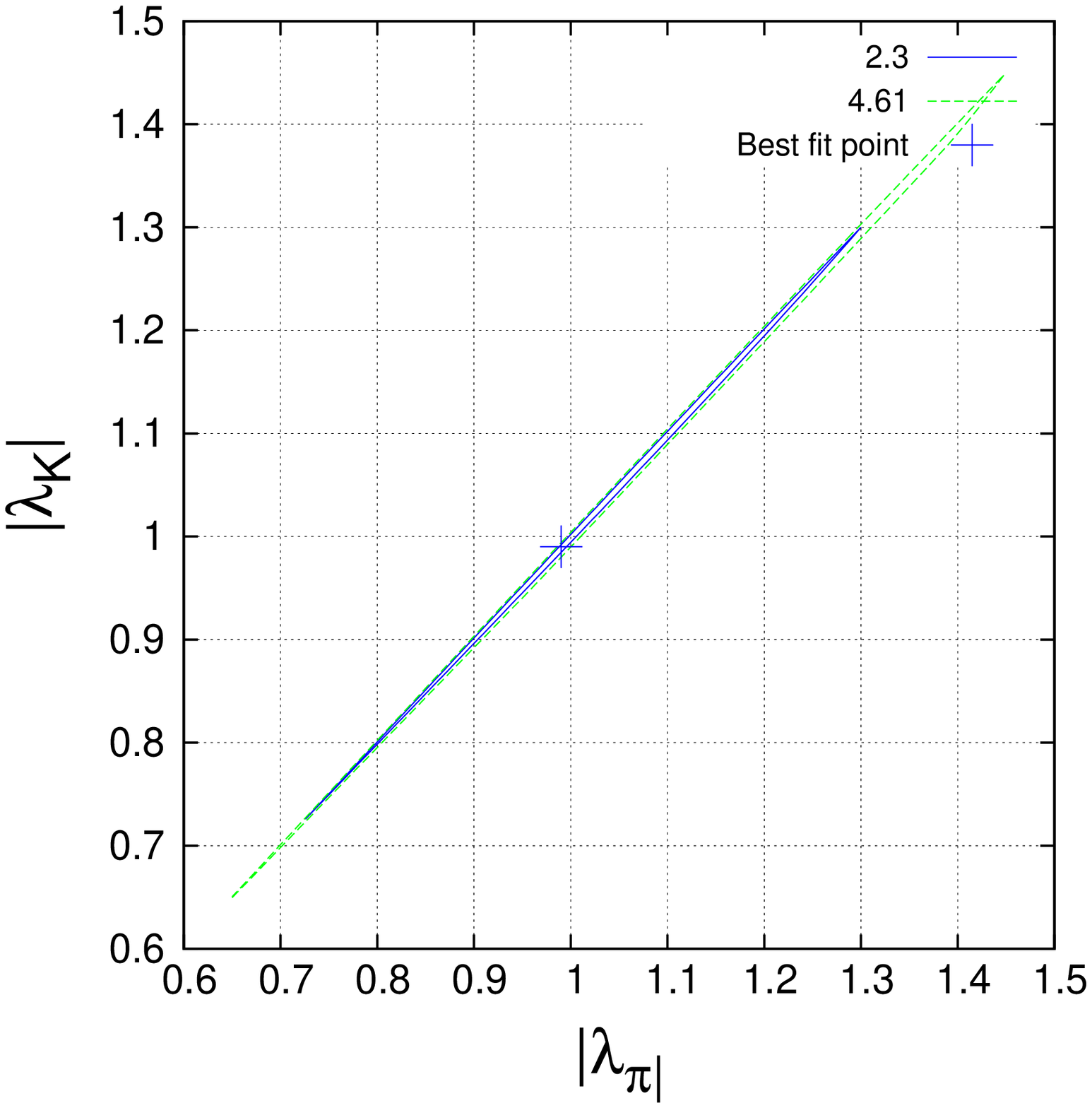}
\includegraphics[width=0.4\textwidth,height=0.49\textwidth,angle=-90]
{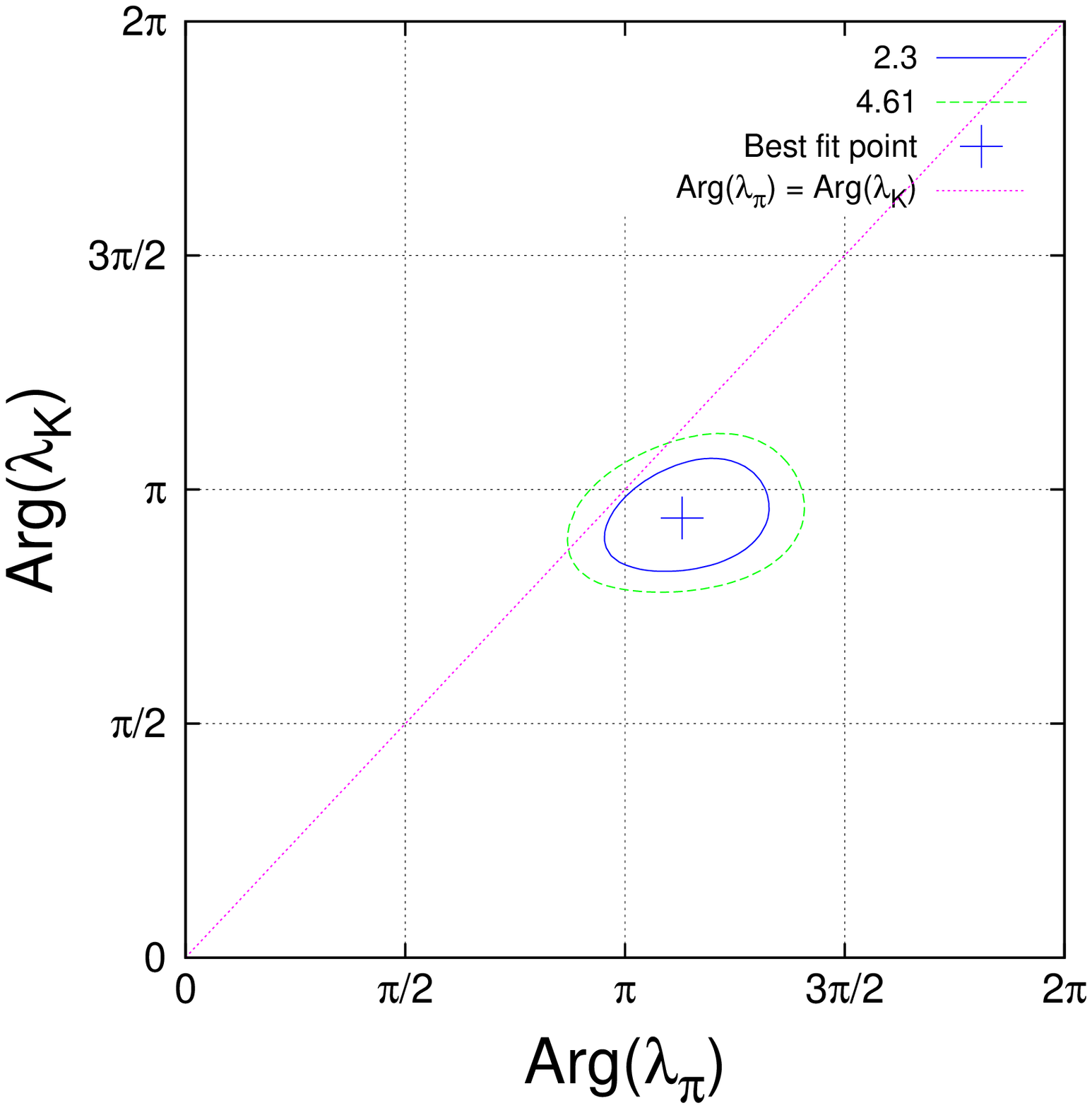}

\caption{Constraints on the magnitudes and phases of $\lambda_\pi$
and $\lambda_K$.}
\label{fig:lambda}
\end{figure}

Let us now explore the extent of CP violation through the interference
of mixing and decay. This may be parametrized through the imaginary
part of $\lambda_\pi \equiv (q/p) R_\pi$ and $\lambda_K \equiv (q/p) R_K$.
While the phases of $q/p, R_\pi$ and $R_K$ shown above are 
convention dependent, the phases of $\lambda_\pi$ and $\lambda_K$
are physical quantities, independent of the phase convention used.
We show the allowed ranges of the magnitudes and phases of $\lambda_\pi$
and $\lambda_K$ in Fig.~\ref{fig:lambda}.

\begin{itemize}
\item The magnitudes of $\lambda_\pi$ and $\lambda_K$ are
highly correlated: $|\lambda_{\pi}| \approx |\lambda_K|$. 
This is expected, since $|\lambda_f|=|(q/p) R_f|$,
wherein the quantity $|q/p|$ is common to both the decay modes, and
$|R_\pi|$ and $|R_K|$ are both very close to unity.

\item On the other hand, different phases for $\lambda_\pi$ and $\lambda_K$ 
seem to be preferred. This leads to a difference in the values of 
$\lambda_\pi$ and $\lambda_K$, and hence different 
values of $\acpindir$ in the two modes.
\end{itemize}

\begin{figure}

\includegraphics[width=0.4\textwidth,height=0.49\textwidth,angle=-90]
{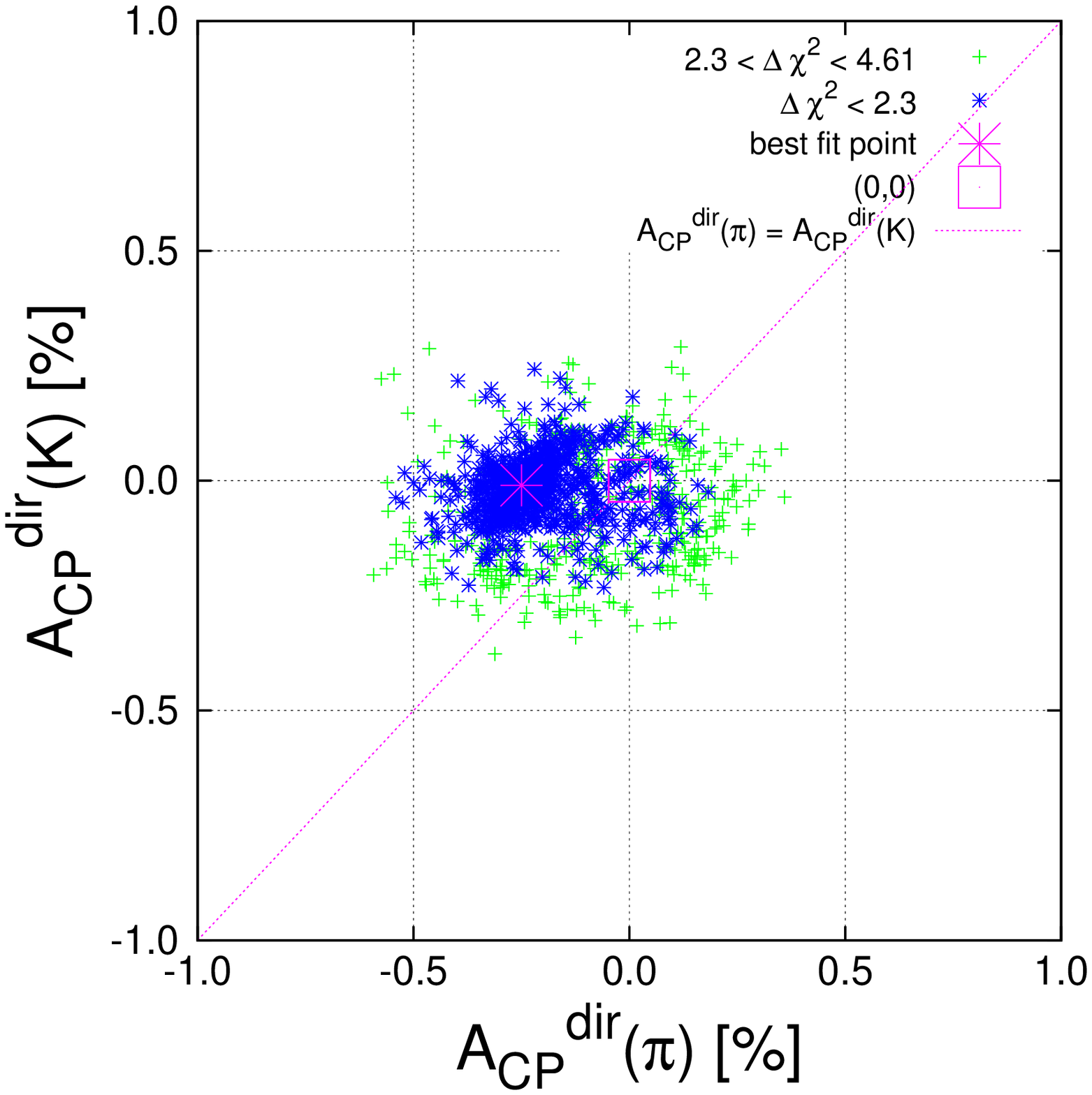}
\includegraphics[width=0.4\textwidth,height=0.49\textwidth,angle=-90]
{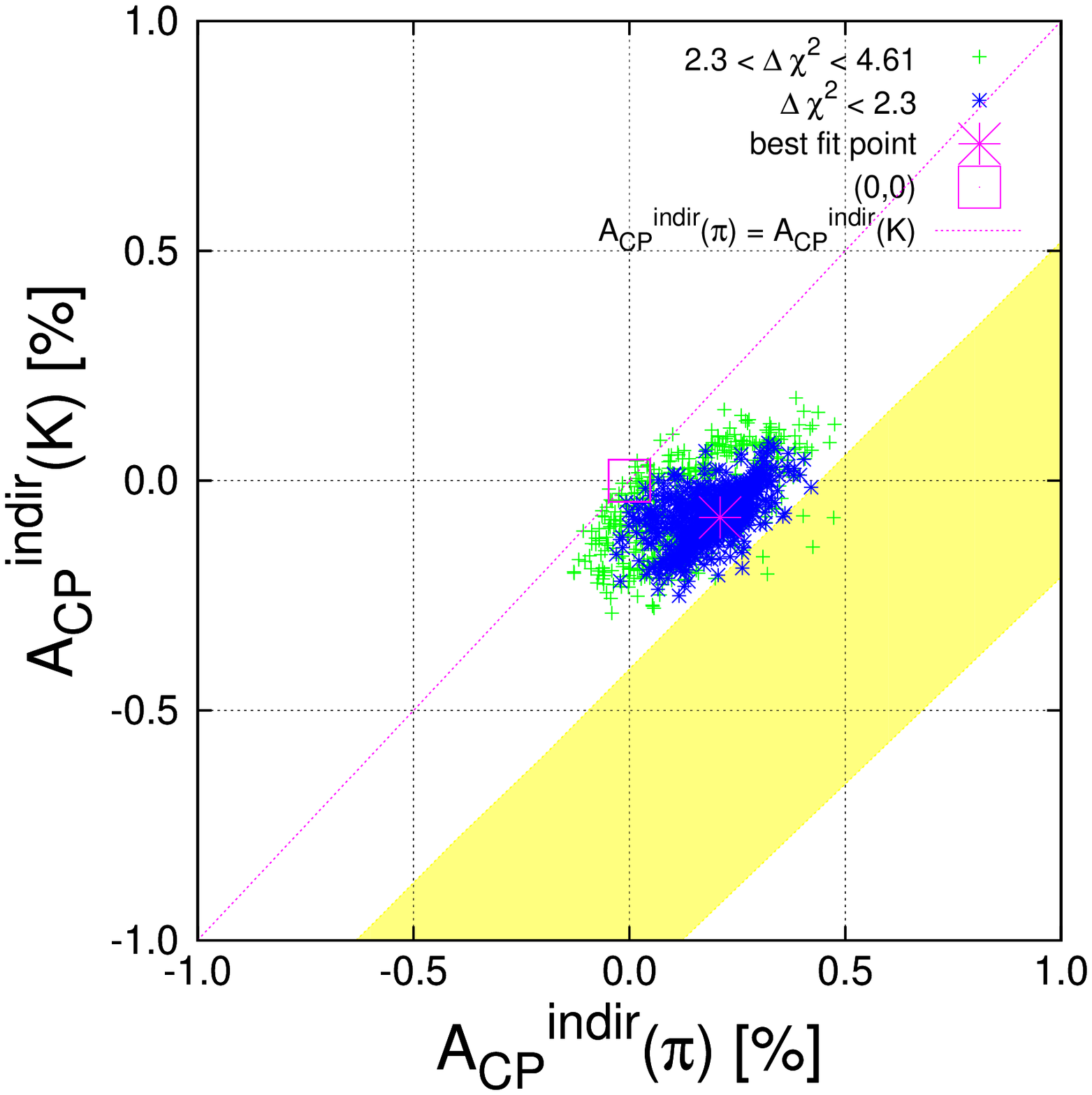}
\caption{Constraints on direct and indirect asymmetries in $D \to \pi\pi$
and $D \to KK$ from the data in
Tables~\ref{tab:acp}, \ref{tab:ycp} and \ref{tab:ddbar}.
The yellow (gray) band in the plot 
on the right corresponds to the values that will reconcile the $\Delta \acp$ 
measurements through the pion-tagged and muon-tagged samples at the LHCb
to within $1\sigma$.}
\label{fig:scatter}
\end{figure}

With the allowed ranges of parameters as determined above, we
present the allowed values of $\acpdir$ and $\acpindir$ in the
form of a scatter plot in Fig.~\ref{fig:scatter}. We can observe 
the following:
\begin{itemize}
\item At the best-fit point,  we have
\begin{center}
\begin{tabular}{ll}
$\acpdir(\pi) = -0.0024$ , &
$\acpindir(\pi) = 0.0021$ , \\ 
$\acpdir(K) = -0.0001$ , &
$\acpindir(K) = -0.0008$ , \\
\end{tabular}
\end{center}
so that the data favors different indirect CP asymmetries in these
two modes. 

\item While the direct CP violation $\acpdir$ in the $\pp$ mode is
restricted to be less than a per cent, that in the $\kk$ mode is
restricted even more severely, to be less than half a per cent.
These allowed values are still much larger as compared to the 
SM expectations. 

\item Indirect CP violation to the extent of half a per cent is still
allowed for $\pp$, while in the case of $\kk$ it can be maximum 
up to a quarter of a per cent. More importantly, different values
of $\acpindir$ in these modes are highly preferred by the data.

\item The shaded band shows that region in which the apparent discrepancy 
between the $\Delta \acp$ measurements from the pion-tagged and 
muon-tagged sample is resolved to within $1\sigma$.
As expected, the resolution favors significantly different values of
$\acpindir$, which is consistent with the results of our fit,
almost to $\sim 1 \sigma$.
Referring back to Eq.~(\ref{eq:delta-delta-acp}),
the large coefficient of the $[\acpindir(K)-\acpindir(\pi)]$
(see Table~\ref{tab:times}) allows such an explanation of
the apparent discrepancy through a moderate difference in the
indirect CP asymmetries in the two decay modes.

\end{itemize}

\begin{figure}
\includegraphics[width=0.4\textwidth,height=0.49\textwidth,angle=-90]
{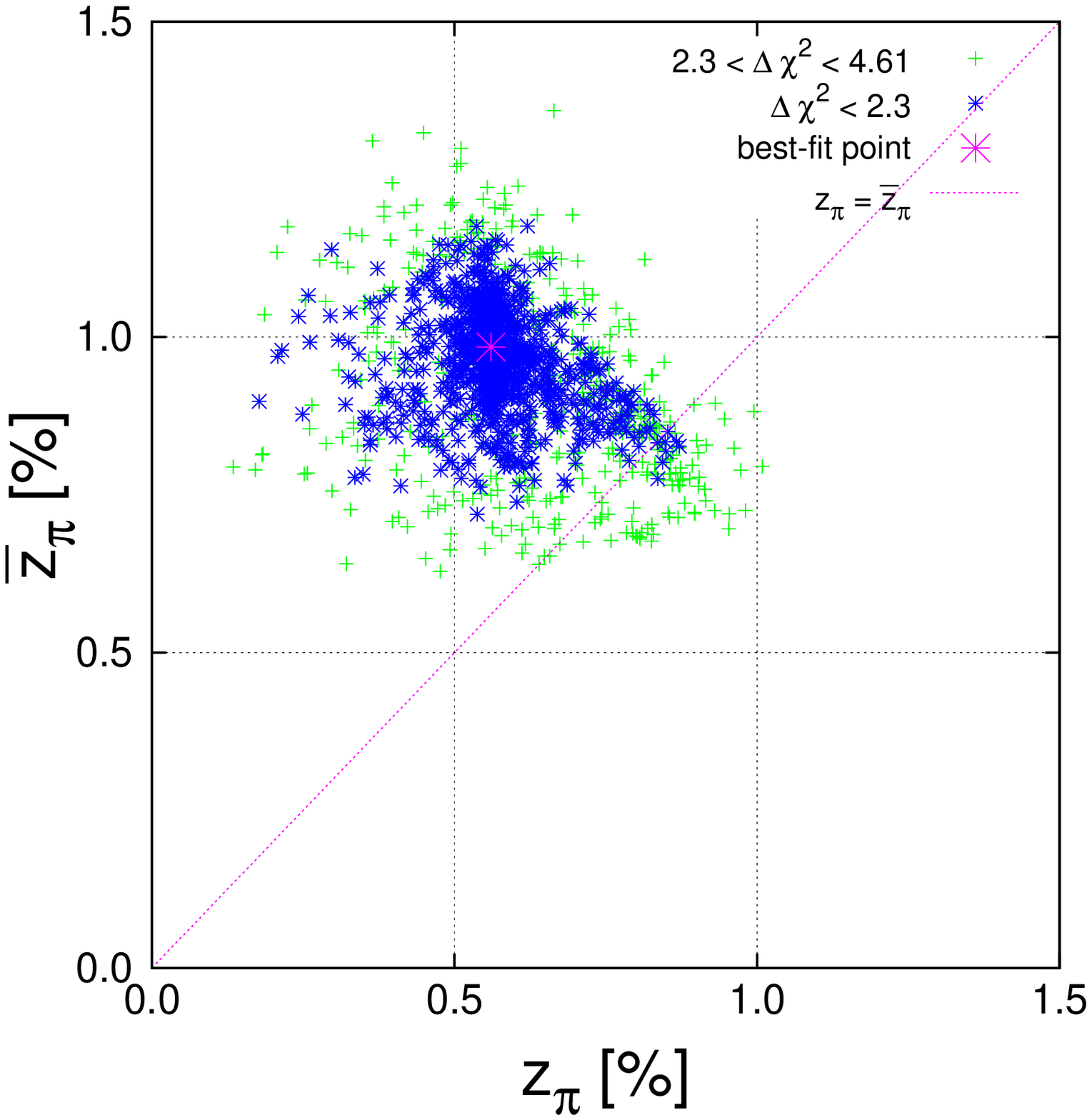}
\includegraphics[width=0.4\textwidth,height=0.49\textwidth,angle=-90]
{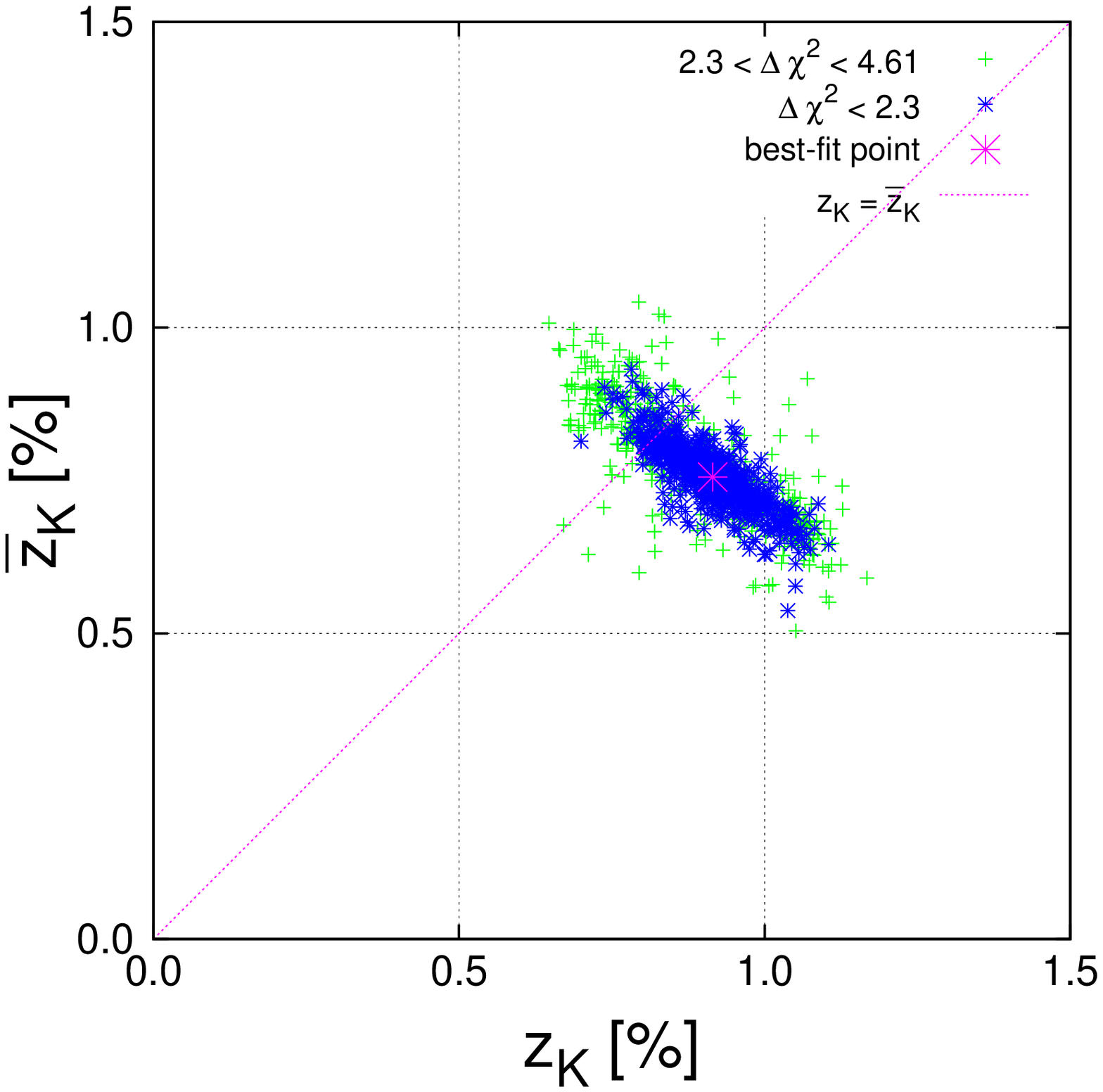}
\caption{Bounds on the parameters $z_\pi$,  $\bar{z}_\pi$,  
$z_K$,  and $\bar{z}_K$, from the measurements of 
$\agamma(\pi), \agamma(K), \ycp(\pi)$ and $\ycp(K)$.}
\label{fig:zzbar}
\end{figure}

Before continuing, we also present the information on the quantities
\barr
z_\pi  \equiv x ~{\rm Im}(\lambda_\pi) -y ~{\rm Re}(\lambda_\pi) \; , & 
 \quad & 
\bar{z}_\pi  \equiv  x ~{\rm Im}(\lambda_\pi^{-1}) 
-y ~{\rm Re}(\lambda_\pi^{-1}) \; ,
\nonumber \\ 
z_K \equiv  x ~{\rm Im}(\lambda_K) -y ~{\rm Re}(\lambda_K) \; ,
& \quad &  
\bar{z}_K  \equiv x ~{\rm Im}(\lambda_K^{-1}) 
-y ~{\rm Re}(\lambda_K^{-1}) \; , 
\earr
obtained from the measurements of the quantities $\agamma$ and $\ycp$.
It may be observed from Fig.~\ref{fig:zzbar} that the favored regions
in the $(z_\pi - \bar{z}_{\pi})$ and $(z_K - \bar{z}_K)$ parameter
space are quite different; they have only a small overlap.
This is in consonance with our overall observation that the data 
indicate unequal amount of CP violation in $\pp$ and $\kk$ modes.
It is therefore important that the measurements of 
$\agamma(\pi), \agamma(K), \ycp(\pi)$ and $\ycp(K)$
be available separately, without averaging over the $\pp$ and
$\kk$ modes,
and analyzed without the assumption of their equality.


\section{Feasibility and implications of nonuniversal $\acpindir$}
\label{implications}

Our analysis as such does not depend on whether we
have only SM, or whether NP is present in addition.
However within the SM, even given the uncertainties due to
long-distance contributions, it is very difficult to get CP violation
of the order of $1\%$ or larger. Indeed, if $|q/p|$ actually
differs substantially from unity, it will need a large phase 
difference between $M_{12}$ and $\Gamma_{12}$, which does not seem
possible in the SM. Also, getting significantly different phases
for the quantities $R_\pi$ and $R_K$, as indicated by the data,
is not something the SM can do. We therefore interpret our results
in terms of the demands they make on NP models, and the observations
required for identifying such NP. To compare our results with 
previous analyses, here we present our arguments in terms of the language 
and notation used in 
Refs.~\cite{pdg-2012,grossman-kagan-nir,kagan-sokoloff}.

In the notation of the Particle Data Group \cite{pdg-2012}, the
decay amplitudes $A_f$ and $\bar{A}_f$ [see Eq.~(\ref{eq:af-abarf})]
are written as
\begin{eqnarray}
A_f & = & A_f^T ~ e^{+ i \phi_f^T} \left[ 1 + r_f ~ e^{i (\delta_f + \phi_f)} 
\right] \; , \nonumber \\
\bar{A}_f & = & A_f^T ~ e^{- i \phi_f^T} \left[ 1 + r_f ~ e^{i (\delta_f - \phi_f)} 
\right] \; , 
\end{eqnarray}
where $A_f^T$ is the leading tree amplitude with its corresponding phase
$\phi_f^T$, while $r_f$ is the ratio of the subleading to the leading
amplitude, with the corresponding strong and weak phase differences  
$\delta_f$ and $\phi_f$, respectively. This notation may be matched to
ours by using
\beq
R_f = \frac{\bar{A}_f}{A_f} = 
e^{-2 i \phi_f^T} \left[ \frac{1 + r_f ~ e^{i (\delta_f - \phi_f)}}{
1 + r_f ~ e^{i (\delta_f +\phi_f)}} \right]  \;.
\eeq
The approximation $r_f \ll1$ used in \cite{pdg-2012,grossman-kagan-nir}
leads to universal indirect CP violation ($= a^m + a^i$) that is independent
of $r_f, \delta_f$ and $\phi_f$. As a result, the quantities 
$\agamma$ and $\ycp$ are also universal, i.e. identical for the
$\pp$ and $\kk$ modes.   

Although the approximation $r_f \ll 1$ is valid in the SM, it is not 
guaranteed to be true in the presence of NP.
Higher-order terms in $r_f$ give nonuniversal contributions
to $\acpindir$ through different values for $\lambda_f$ 
\cite{kagan-sokoloff}. The value of $r_f$ is bounded by the CP violation
measurements themselves: since $\acpdir(f) \sim {\cal O}(0.01)$,
we have $|R_f| \sim 1+ {\cal O}(0.01)$, and hence 
$r_f \sin \delta_f \sin \phi_f \sim {\cal O}(0.01)$, for both the decay modes.  
On the other hand, in order to have a significant nonuniversality in
$\acpindir$ through $\lambda_\pi \neq \lambda_K$,
we need ${\rm Arg}(\lambda_\pi) \neq {\rm Arg}(\lambda_K)$, since 
our fit suggests equal magnitudes for these two quantities 
(see Fig.~\ref{fig:lambda}.
Using the results in \cite{kagan-sokoloff},
the relevant nonuniversality may be expressed in terms of
\beq
{\rm Arg}(\lambda_K) - {\rm Arg}(\lambda_\pi) \approx
2 r_\pi \cos \delta_\pi \sin \phi_\pi - 2 r_K \cos \delta_K \sin \phi_K \; ,
\eeq
so that at least one of the two quantities $r_f \cos \delta_f \sin \phi_f$
should be significantly large, i.e. ${\cal O}(0.1-1)$.
For both the above constraints to be satisfied simultaneously,
we need $\sin \phi_f \sim {\cal O}(1)$, 
$r_f \sin \delta_f \sim {\cal O}(0.01)$, and  
$r_f \cos \delta_f \sim {\cal O}(0.1-1)$,
for at least one of the final states $\pp$ and $\kk$.

The above argument suggests that for the condition implied by our
best-fit point to be met, one needs an enhanced subleading amplitude with
a large relative weak phase and a small relative small phase
compared to the leading term, for at least one of the two final states.
While the first two conditions may be generically satisfied by a NP
model that is not too constrained from other measurements, the smallness
of the relative strong phase may seem to be rather fine-tuned, since
the NP operators typically differ from the leading ones in their color
and chirality structure \cite{kagan-sokoloff}.
Explicit calculations or direct measurements of these strong 
phase differences are not available; recent analysis of the $\pi\pi$ 
scattering data in the context of SM \cite{franco-silvestrini}
indicates large strong phases, but note that what is needed here 
is a small {\it difference} in the strong phases of the leading and 
subleading contributions. 
On the other hand, the magnitude and phase of the NP contribution
should conspire such that $|R_f| \approx 1.00$ to within $1\%$.
The NP scenario required here is thus rather fine-tuned, but is
not ruled out, and hence should not be ignored. Further, note that if the
difference between the LHCb measurements with the pion-tagged and 
muon-tagged samples is real, this is perhaps the only mechanism that can 
account for it. In fact, the measurement of nonuniversality of 
$\acpindir$ itself would give direct information on the smallness or
largeness of this phase that appears in all the CP-violating observables.

The NP scenario described above, which gives large nonuniversality  
in $\acpindir$, is mandated if the best-fit point obtained from our fit 
indeed turns out to be the right one with future data.
However even if the future data were to indicate smaller nonuniversality
in $\acpindir$ than the best-fit point obtained our fit, our analysis
in Sec.~\ref{analytic} still stays relevant. It gives generalized
expressions for the CP asymmetry $\acp$ and the related quantities
$\agamma, \ycp$ [see Eqs.~(\ref{cpdir}), (\ref{cpindir}), 
(\ref{eq:agamma}), (\ref{eq:ycp})] that are valid even with 
nonuniversal $\acpindir$. The expressions prevalent in the standard 
literature \cite{hfag,pdg-2012} do not take this possibility into account. 

The nonuniversality of $\agamma$ and $\ycp$ has been explicitly 
calculated in \cite{gersabeck-12}, albeit with different expansion 
parameters than the ones considered here. We believe that our expansion
parameters ($x$ and $y$) are more well motivated since they have been
measured to be small. Moreover, while obtaining the final expression 
for $\Delta \acp$, Ref.~\cite{gersabeck-12} used the assumption of universality 
of the phase of $\lambda_f$, thus restricting its domain of validity.

\section{Summary and conclusion}
\label{concl}

The recent measurements of the CP asymmetries in $D/\bar{D} \to
\pp, \kk$ modes, and the difference in the CP asymmetries
in these modes (the so-called difference CP asymmetry) have 
yielded values differing from the SM expectations.
Moreover, the difference CP asymmetries measured at the LHCb through 
the pion-tagged and the muon-tagged samples differ substantially.
We examine these data in a model-independent manner to discern the 
nature of CP violation involved therein and to find a resolution
for the above discrepancy.

By performing a fit to the data on $\ddbar$ mixing, CP asymmetries
in $D/\bar{D} \to \pp, \kk$ modes, as well as the related quantities
$\agamma$ and  $\ycp$ in these channels, we find the following:
(i) The CP violation through decay-only in both the modes is restricted 
to be less than ${\cal O}(0.5\%)$, although this limit still allows values 
much larger than those permitted by the SM. 
(ii) The CP violation through mixing-only, on the other hand, can be 
quite large --- the value of $|q/p|$ can differ from unity by 
$\sim {\cal O}(10\%)$.
(iii) The CP violation through the mixing-decay interference may play
an important role in the $D/\bar{D} \to \pp, \kk$ modes.
The phases of the quantities $\lambda_\pi$ and $\lambda_K$ tend to
differ substantially.

In the language of direct and indirect CP asymmetries, as used while
presenting the recent experimental data, the direct CP asymmetries
in both, $\pp$ and $\kk$ modes is restricted to be less than 
${\cal O}(0.5\%)$, and these two asymmetries can have different values.
The indirect CP asymmetries in these two modes also can differ
substantially in certain scenarios. 
Indeed we demonstrate that, mathematically speaking,
different direct CP asymmetries imply different indirect CP asymmetries, 
unless there are accidental cancellations. We therefore emphasize that 
taking the indirect CP asymmetries in these two channels to be 
equal is an approximation, one that may be valid, but that should be 
checked with data.

It turns out that the possibility of nonuniversal indirect CP asymmetry 
also allows a partial reconciliation between the
seemingly different difference CP asymmetries measured through the 
pion-tagged and muon-tagged samples at the LHCb.
The analyses for these decay modes in terms of the direct and 
indirect CP asymmetries should therefore be performed without the 
usual assumption of equal indirect CP asymmetries in the two modes.

Our formalism also allows us to express the quantities $\agamma$ 
and $\ycp$ in a symmetric form, $\agamma = (z - \bar{z})/2$ and
$\ycp= (z + \bar{z})/2$, without having to assume equal magnitudes as
well as phases for $A(\bar{D} \to f)/A(D \to f)$ for the
$\pp$ and $\kk$ channels. This also indicates that the data on these
quantities in the $\pp$ and $\kk$ channels should be presented
separately, since these quantities can be different in these two 
modes and an averaging might lose information critical for
ascertaining the presence and nature of any NP present.

A significant nonuniversality in $\acpindir$ would require
the subdominant amplitude in $D \to f$ decay to be comparable in
magnitude to the dominant one, as well as a small strong relative 
phase  and a large weak relative phase between these two amplitudes.
This scenario appears fine-tuned, given the theoretical expectation 
of large relative strong phases; however, there is no direct calculation 
or measurement of the strong phase, so it is not ruled out. 
Therefore, it is important to take into account the possibility of such
a NP that could give rise to significant nonuniversality in $\acpindir$,
$\agamma$, and $\ycp$.  The generalized analysis and expressions 
presented in this paper then need to be used instead of the ones 
in standard literature that assume universality of $\acpindir$. 
Indeed, such a generalized analysis will be useful in measuring the
extent of the nonuniversality itself, and testing of the theoretical
expectation of a large strong relative phase.

Since using the generalized analysis (by removing just one assumption) 
offers the possibility of explaining the current data as well as
probing NP signals with future data, this opportunity should not be missed. 
This implies that the averaging of $\agamma, \ycp$ values in $\pp,\kk$ 
modes at the B factories, as well as the averaging of $\Delta \acp$ in 
pion-tagged and muon-tagged modes in LHCb, should be avoided. With large 
amounts of data around the corner from the LHC upgrade and the super-B 
factory, statistics will cease to be the limiting factor, and one can probe
possible NP in the CP violation in $D$ decays in a clean way.


\section*{Note Added}

\begin{figure}

\includegraphics[width=0.4\textwidth,height=0.49\textwidth,angle=-90]
{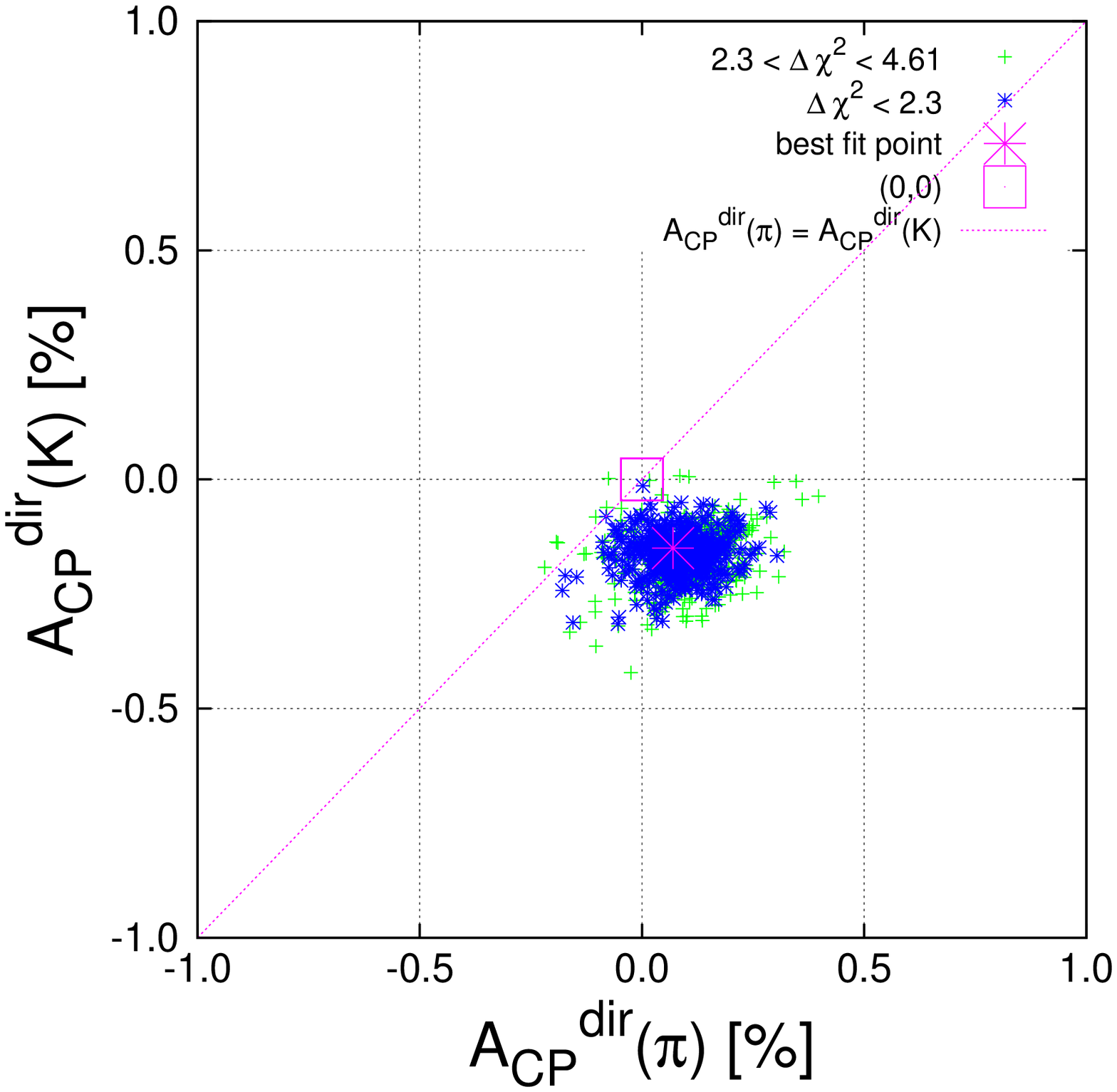}
\includegraphics[width=0.4\textwidth,height=0.49\textwidth,angle=-90]
{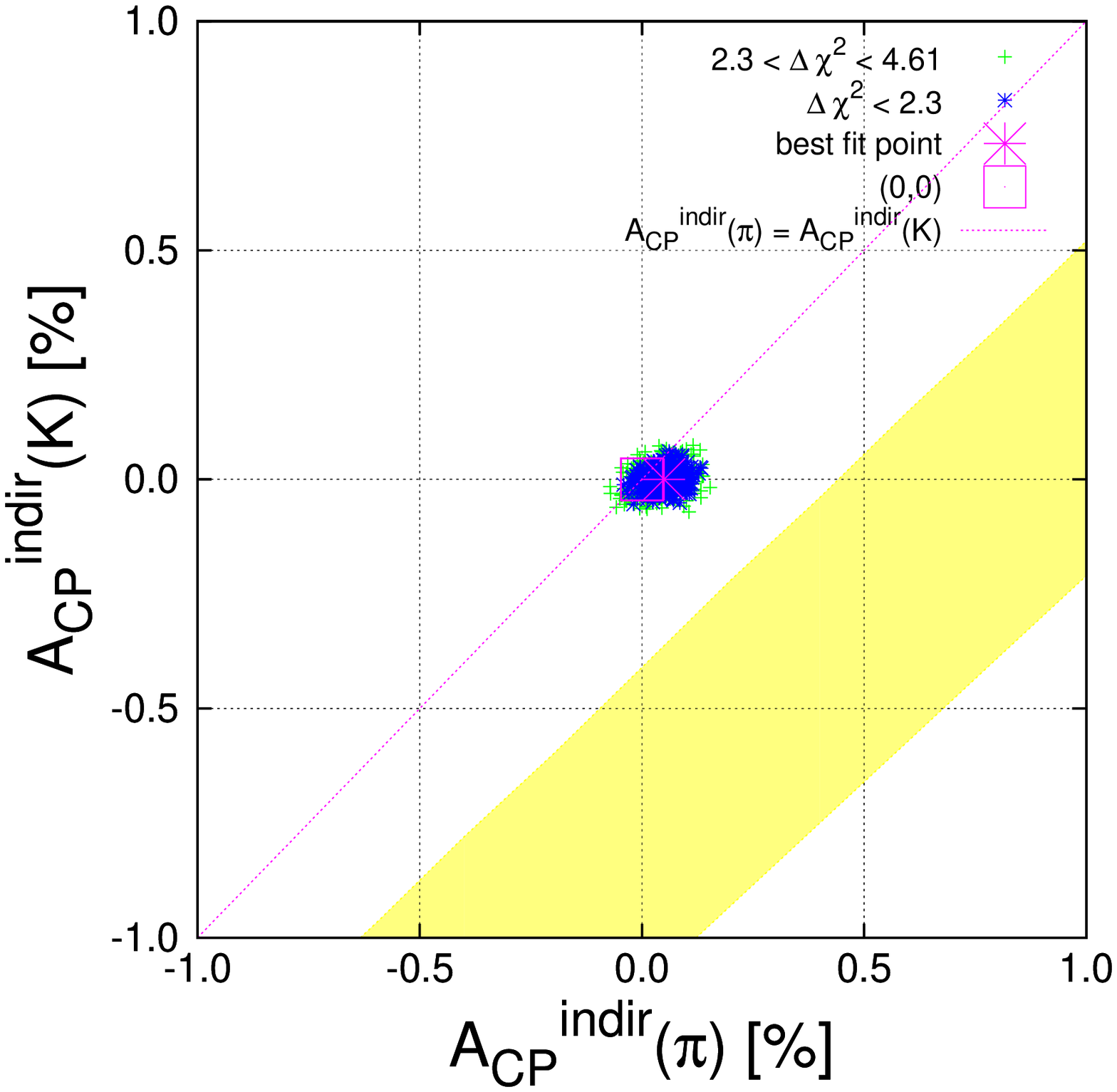}
\caption{Constraints on direct and indirect asymmetries in $D \to \pi\pi$
and $D \to KK$ from the current data, with the new LHCb 2013 
\cite{lhcb-agamma-2013} results added. The yellow (gray) band in the plot 
on the right corresponds to the values that will reconcile the $\Delta \acp$ 
measurements through the pion-tagged and muon-tagged samples at the LHCb
to within $1\sigma$.
}
\label{fig:scatter_rev}
\end{figure}

While this article was under review, the LHCb Collaboration
announced undated results on $\agamma$ \cite{lhcb-agamma-2013}:
$\agamma(K) = (-0.035 \pm 0.062 \pm 0.012)\%$, 
$\agamma(\pi) = (0.033 \pm 0.106 \pm 0.014)\%$.
Adding these measurements to our fit, we find that 
(see Fig.~\ref{fig:scatter_rev}) the ability 
of nonuniversal-$\acpindir$ to help reconcile the $\Delta \acp$ 
measurements through the pion-tagged and muon-tagged samples at the LHCb
is rather restricted by the new data. However the last word 
on the CP violation in $D$ meson system is yet to be written
\cite{lenz-charm-2013}, and a complete analysis should take into 
account the possibility of different indirect CP asymmetries in the 
$\pi^+\pi^-$ and $K^+ K^-$ channels.

\section*{Acknowledgments}

We would like to thank Marco Gersabeck, Alex Kagan, Yossi Nir, 
Luca Silvestrini and Rahul Sinha for useful comments and scintillating
discussions on the first version of the manuscript.
DG would like to thank Satoshi Mishima, Ayan Paul and Luca Silvestrini 
for useful discussions. 
The research by D. G. leading to these results has received funding from 
the European Research Council under the
European Union's Seventh Framework Programme (FP/2007-2013)/ERC Grant 
No. 279972. D. G. also acknowledges the hospitality of the SLAC 
Theoretical Physics Group where this work was completed. 

\end{document}